\documentclass[submission, Phys]{SciPost}

\usepackage{graphicx}
\usepackage{dcolumn}
\usepackage{bm}
\usepackage{epstopdf}
\usepackage{float}
\usepackage{amsmath}
\usepackage{color}
\usepackage{comment}

\hypersetup{
    colorlinks,
    linkcolor={red!50!black},
    citecolor={blue!50!black},
    urlcolor={blue!80!black}
}

\usepackage[bitstream-charter]{mathdesign}
\DeclareSymbolFont{usualmathcal}{OMS}{cmsy}{m}{n}
\DeclareSymbolFontAlphabet{\mathcal}{usualmathcal}

\begin{document}

\begin{center}{\Large \textbf{
Polarons and bipolarons in a two-dimensional square lattice\\
}}\end{center}

\begin{center}
Shanshan Ding\textsuperscript{1},
G. A. Dom\'inguez-Castro\textsuperscript{2,3},
Aleksi Julku\textsuperscript{1},
Arturo Camacho-Guardian\textsuperscript{2} and
Georg M.\ Bruun\textsuperscript{1,4$\star$}
\end{center}

\begin{center}
{\bf 1} Center for Complex Quantum Systems, Department of Physics and Astronomy, Aarhus University, Ny Munkegade, DK-8000 Aarhus C, Denmark
\\
{\bf 2} Instituto de F\'isica, Universidad Nacional Aut\'onoma de M\'exico, Apartado Postal 20-364, Ciudad de M\'exico C.P. 01000, Mexico
\\
{\bf 3} Institut f\"ur Theoretische Physik, Leibniz Universit\"at Hannover, Appelstrasse 2, DE-30167 Hannover, Germany
\\
{\bf 4} Shenzhen Institute for Quantum Science and Engineering and Department of Physics, Southern University of Science and Technology, Shenzhen 518055, China
\\
${}^\star$ {\small \sf bruungmb@phys.au.dk}
\end{center}

\begin{center}
\today
\end{center}


\section*{Abstract}
{\bf
Quasiparticles and their  interactions are a key part of our  understanding of quantum many-body systems.
Quantum simulation experiments with cold atoms have in recent years advanced our understanding  of isolated quasiparticles, but  so far
they have provided limited information regarding their interactions and possible bound states. Here, we show how exploring  mobile impurities immersed
in a Bose-Einstein condensate (BEC) in a two-dimensional lattice can address this problem. First, the spectral  properties of individual impurities are examined, and in addition to
the  attractive and repulsive polarons known from continuum gases, we identify a new kind of quasiparticle stable for repulsive boson-impurity interactions.
 The spatial properties of polarons  are calculated showing that there is an increased density of bosons at the site of the impurity both for
 repulsive and attractive interactions.
 We then derive an effective Schr\"odinger equation describing two polarons interacting via the exchange of density oscillations in the BEC, which  takes into
 account strong impurity-boson two-body correlations.  Using this, we show that the attractive nature of the effective interaction between two polarons
 combined with the two-dimensionality of the lattice leads to the formation of bound states -- i.e.\
 bipolarons.  The wave functions of the bipolarons are examined showing that the ground state is symmetric under particle exchange and therefore relevant for bosonic
 impurities, whereas the first excited state is doubly degenerate and odd under particle exchange making it relevant for fermionic impurities.
 Our results show that  quantum gas microscopy in optical lattices is  a promising platform to explore the spatial properties of polarons as well as
 to finally observe the elusive  bipolarons.
}

\vspace{10pt}
\noindent\rule{\textwidth}{1pt}
\tableofcontents\thispagestyle{fancy}
\noindent\rule{\textwidth}{1pt}
\vspace{10pt}

\section{Introduction}
\label{sec:intro}
In seminal works, Landau, Pekar, and Fr\"ohlich~\cite{landau1948effective,Frohlich1954} demonstrated
that an electron moving through a dielectric distorts the surrounding crystal lattice so that it is dressed by phonons, which leads to the formation of a particle-like object coined a quasiparticle. Landau then realized that this effect is more general and that many quantum systems can be described in terms of interacting
quasiparticles with properties that may be quite different from the underlying bare particles~\cite{baym2008landau,pines2018theory}. This insight stands out as a highlight in theoretical physics, since it provides a simple yet accurate description of a wide range of natural systems including electrons in solids,  liquid helium mixtures, nuclear matter,
elementary particles, and even biological systems \cite{baym2008landau,pines2018theory,ring2004nuclear,Weinberg1995,Sarovar:2010aa}.

The realisation of quasiparticles in quantum degenerate atomic gases has allowed a systematic investigation
of their properties~\cite{Nascimbene2009,Schirotzek2009,Kohstall:2012aa,Koschorreck:2012aa,Scazza2017,Jorgensen2016,Hu2016,Yan2020}.
In these gases, the quasiparticle, called a polaron, consists of an impurity atom interacting with a surrounding atomic gas and
we now have an accurate description  of individual  polarons formed in fermionic gases~\cite{Massignan_2014,Scazza2022},
 whereas open questions  remain regarding polarons in Bose-Einstein condensates (BECs) for strong interactions. So far, these experiments have however not revealed much information regarding the
 interactions between the quasiparticles, even though they have been predicted to exist both for polarons in Fermi~\cite{Heiselberg2000,Hu2018,Mora2010,Tajima2021,Muir2022,Fujii2022} and
 Bose gases~\cite{Fujii2022,Yu2012,Naidon2018,Dehkharghani2018,Camacho-Guardian2018}. This is an important open question, since such interactions are an inherent property of quasiparticles
  as they affect each other by modulating the surrounding medium. Indeed, interactions between quasiparticles are
  instrumental for understanding their thermodynamic as well as dynamical properties~\cite{baym2008landau,pines2018theory}. A
  particularly striking effect is that they can give rise to bound states. This
 is the origin of conventional superconductors where the  interaction is mediated by phonons~\cite{Schrieffer1983}, and
 a quasiparticle interaction mediated by spin fluctuations is conjectured to be the mechanism behind
 high temperature superconductivity~\cite{SCALAPINO1995329,Lee2006}. In the case of polarons, such
 bound states are called bipolarons, and suggested as a mechanism for charge transport in polymers~\cite{Bredas:1985vc} as well as
magnetoresistance in  organic materials~\cite{Bobbert2007}. Bipolarons have also been predicted to exist in atomic BECs~\cite{camacho2018bipolarons}
but have so far not been observed.

Ultracold atoms in optical lattices represent a  powerful quantum simulation platform for many-body physics since they realize
the Hubbard model essentially perfectly~\cite{bloch2008many}. Moreover,
 the ability to take pictures of individual atoms using quantum gas microscopy provides  detailed spatial information of the quantum states
 that compliments the usual information obtained from spectroscopy~\cite{Bakr:2009aa,Sherson:2010aa}. This  has led to a range of
experimental  breakthroughs regarding quantum magnetism, topological matter, quantum phase transitions,  and non-equilibrium physics~\cite{Bloch:2012aa,Gross2017}.
Impurities in a one-dimensional (1D) lattice containing a BEC have been considered~\cite{Privitera2010,Massel_2013,Keiler_2020,Sarkar_2020},
and the effects of the Mott-insulator to superfluid transition of bosons in a 2D square lattice on the polaron have  been analyzed for weak boson-impurity interactions~\cite{Colussi2022}.

Here, we show that  impurity atoms in a 2D optical lattice containing an atomic BEC  represent a
promising setup to observe the spatial properties of  polarons and the formation of bipolarons for the first time. We first analyze the properties of single polarons and demonstrate that as a consequence of the lattice, a new undamped quasiparticle branch that has no analogue in continuum systems emerges for repulsive interactions.
The spatial correlations between the impurity  and the surrounding bosons are explored and we show that the boson density is significantly
 enhanced at the impurity site both for attractive and repulsive interactions. Having analysed individual polarons, we then
 derive an effective  non-local Schr\"odinger equation describing the dynamics of two interacting polarons in the BEC. The interaction between the polarons is
 mediated by Bogoliubov modes in the BEC, and we show that it is attractive and can be strong enough to support bound states.
While the wave function of the ground state bipolaron has an $s$-wave character and is symmetric under particle exchange, the first excited doubly degenerate bipolaron
states have a $p$-wave character and are odd under particle exchange, making them relevant to bosonic or fermionic impurities respectively.
We argue that the bipolarons can be  observed via the spatial correlations between two impurities using quantum gas microscopy available
in optical lattices.

\section{System}
\begin{figure}[ht]
\centering
\includegraphics[width=0.7\textwidth]{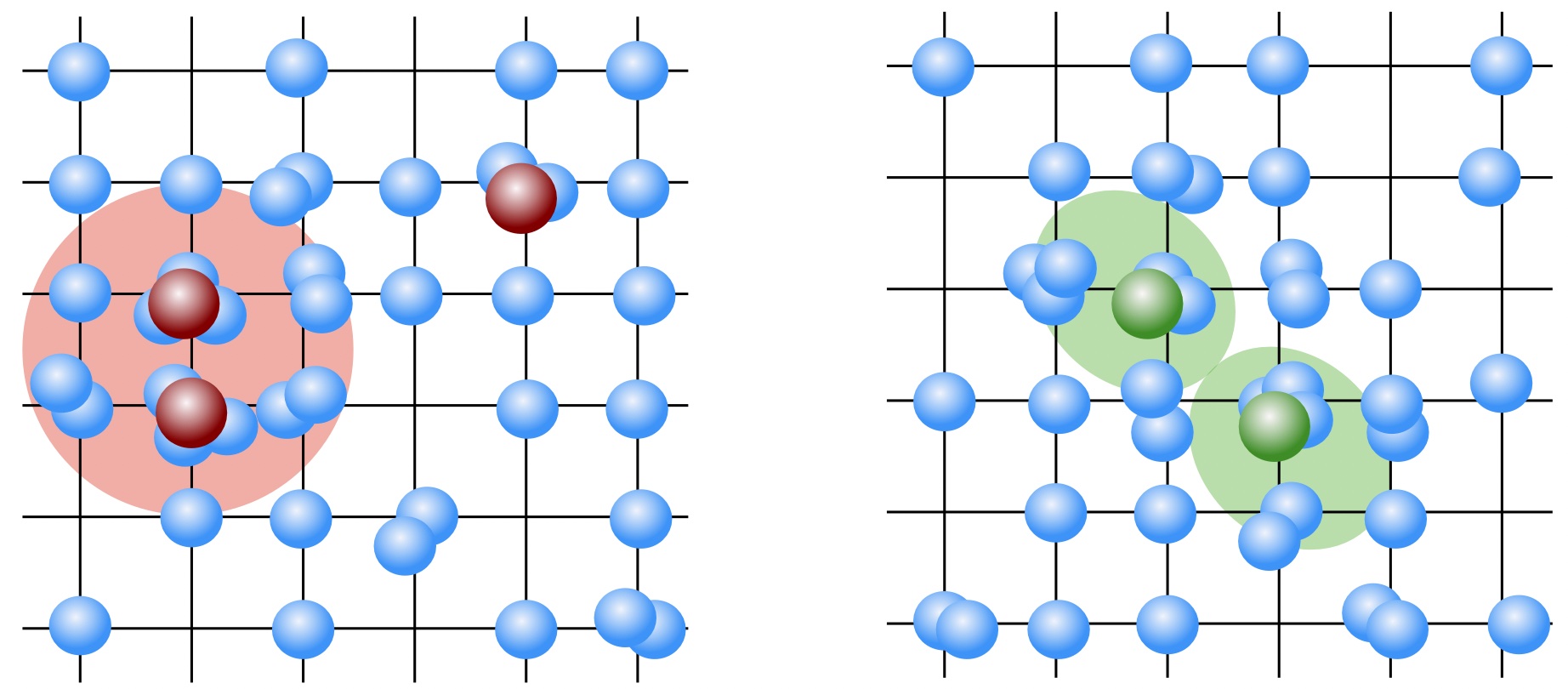}%
\caption{Illustration of the system considered. Mobile bosonic or fermionic impurities (red and green balls respectively)
 in a square lattice containing a BEC (blue balls). Interactions between the
BEC and the impurities led to the formation of polarons, which in turn can form bound states, i.e.\  bipolarons, due to an induced interaction mediated by density ripples in the
BEC. When the impurities are bosonic, the ground state bipolaron has $C_4$ ``$s$-wave'' symmetry whereas it has $C_2$ ``$p$-wave''  symmetry for fermionic impurities.
}
\label{Fig-setup}
\end{figure}

We consider  mobile impurities mixed with identical bosons in a 2D square lattice,  see Fig.~\ref{Fig-setup}.
The Hamiltonian is
\begin{equation}\begin{aligned}
\hat H=&-t_B\sum_{ \langle {\mathbf i},{\mathbf j}\rangle}\hat{b}_{\mathbf i}^\dagger \hat{b}_{\mathbf j}+\frac{U_B}{2}\sum_{{\mathbf i}}\hat{b}_{\mathbf i}^\dagger \hat{b}_{\mathbf i}^\dagger \hat{b}_{\mathbf i} \hat{b}_{\mathbf i}
-\mu_B \sum_{\mathbf i} \hat{b}_{\mathbf i}^\dagger \hat{b}_{\mathbf i}\\
&-t_I\sum_{\langle {\mathbf i},{\mathbf j} \rangle}\hat{c}_{\mathbf i}^\dagger \hat{c}_{\mathbf j}+\frac{U_I}{2}\sum_{\mathbf i}\hat{c}_{\mathbf i}^\dagger \hat{c}_{\mathbf i}^\dagger \hat{c}_{\mathbf i} \hat{c}_{\mathbf i}
+U_{BI}\sum_{{\mathbf i}}\hat{b}_{\mathbf i}^\dagger \hat{c}_{\mathbf i}^\dagger \hat{c}_{\mathbf i}\hat{b}_{\mathbf i}.
\end{aligned}\label{Eq-H-BEC}
\end{equation}
Here,  operators $\hat{b}_{\mathbf i}$ and  $\hat{c}_{\mathbf i}$ remove a boson and an impurity at lattice site ${\mathbf i}$, the nearest neighbour
hopping matrix element  is $t_B$ for the bosons
 and $t_I$  for the impurities, $U_B>0$ is the on-site boson-boson repulsion, $U_I$ is the on-site impurity-impurity interaction,
 and $U_{BI}$ is the on-site interaction between the bosons and the impurities. 
 The chemical potential of the bosons is $\mu_B$ and we set $\hbar$ and the lattice constant to unity.

The bosons form a BEC with  $n_0$ particles per lattice site, and we assume that the interaction $U_B$ is weak so that
the BEC is accurately described by Bogoliubov theory. This gives the  chemical potential $\mu_B=-4t_B+n_0U_B$ and the
excitation spectrum $E_{\mathbf{k}}=\sqrt{\epsilon_{B\mathbf{k}}^2+2\epsilon_{B\mathbf{k}}n_0U_B}$, where
$\epsilon_{B\mathbf{k}}=-2t_B(\cos k_x+\cos k_y)+4t_B$. Here, $\mathbf{k}$ is the crystal momentum of the excitation within the first
Brillouin zone (BZ) of the lattice. In the rest of the paper, we take for concreteness $t_I=t_B$.

\section{Two-body scattering and bound states}\label{TwobodySec}
We first study the scattering of an impurity and a boson in an empty lattice. For the contact interaction $U_{BI}$,
the scattering matrix  only depends on the total center-of-mass momentum $\mathbf P$ and the energy $\omega$ of the pair and
takes the simple form
\begin{equation}
\mathcal{T}(\mathbf{P}, \omega) = \frac{U_{BI}}{1-U_{BI}\Pi_{\text v}(\mathbf{P}, \omega)}.
\label{ScatteringMatrix}
\end{equation}
Here,
\begin{equation}
\Pi_{\text v}(\mathbf{P}, \omega) = \frac{1}{M}\sum_{\mathbf{k}}\frac{1}{\omega-\epsilon_{B\mathbf{P}/2+\mathbf{k}}-\epsilon_{I\mathbf{P}/2-\mathbf{k}}}
\label{PairpropVac}
\end{equation}
is the pair propagator in an empty lattice with $\epsilon_{I\mathbf{k}}=-2t_I(\cos k_x+\cos k_y)$ and $M$ the number of lattice sites.
 We measure energies with respect to the minimum $-4t_B$ of the
tight-binding band of the bosons in order to facilitate comparisons with the case when a BEC is present and the energies are measured with
respect to $\mu_B$. For an infinite lattice, the pair propagator in Eq.~\eqref{PairpropVac} can be evaluated analytically giving
\begin{equation}
\label{pair}
\Pi_{\text v}(\mathbf 0, \omega)t_B =
\begin{cases}
[\text{sgn}(z)K(|z|)-iK(\sqrt{1-z^2})]/{4\pi}\\
K(|z|^{-1})/4\pi z,
\end{cases}
\end{equation}
where   $K(z)$ is the complete elliptic integral of the first kind,
the top line is for $-1 < z < 1$, and the bottom line is for any other case. We have defined
 $z=(\omega-4t_B)/8t_B$ representing the two-body energy measured in units of $8t_B$. Details of deriving Eq.~\eqref{pair} are given in App.~\ref{PairApp}.
 The non-zero imaginary part of $\Pi(\mathbf 0, \omega)$ for $-4t_B<\omega<12t_B$ comes from the  continuum of  impurity and boson single-particle states
 with zero center of mass (COM) momentum and energies being $\epsilon_{B\mathbf{k}}+\epsilon_{I-\mathbf{k}}$.
 Discrete poles of the scattering matrix on the other hand give the energies of any bound  states consisting of one boson and one impurity in the lattice.

\begin{figure}[!ht]
\centering
\includegraphics[width=0.7\textwidth]{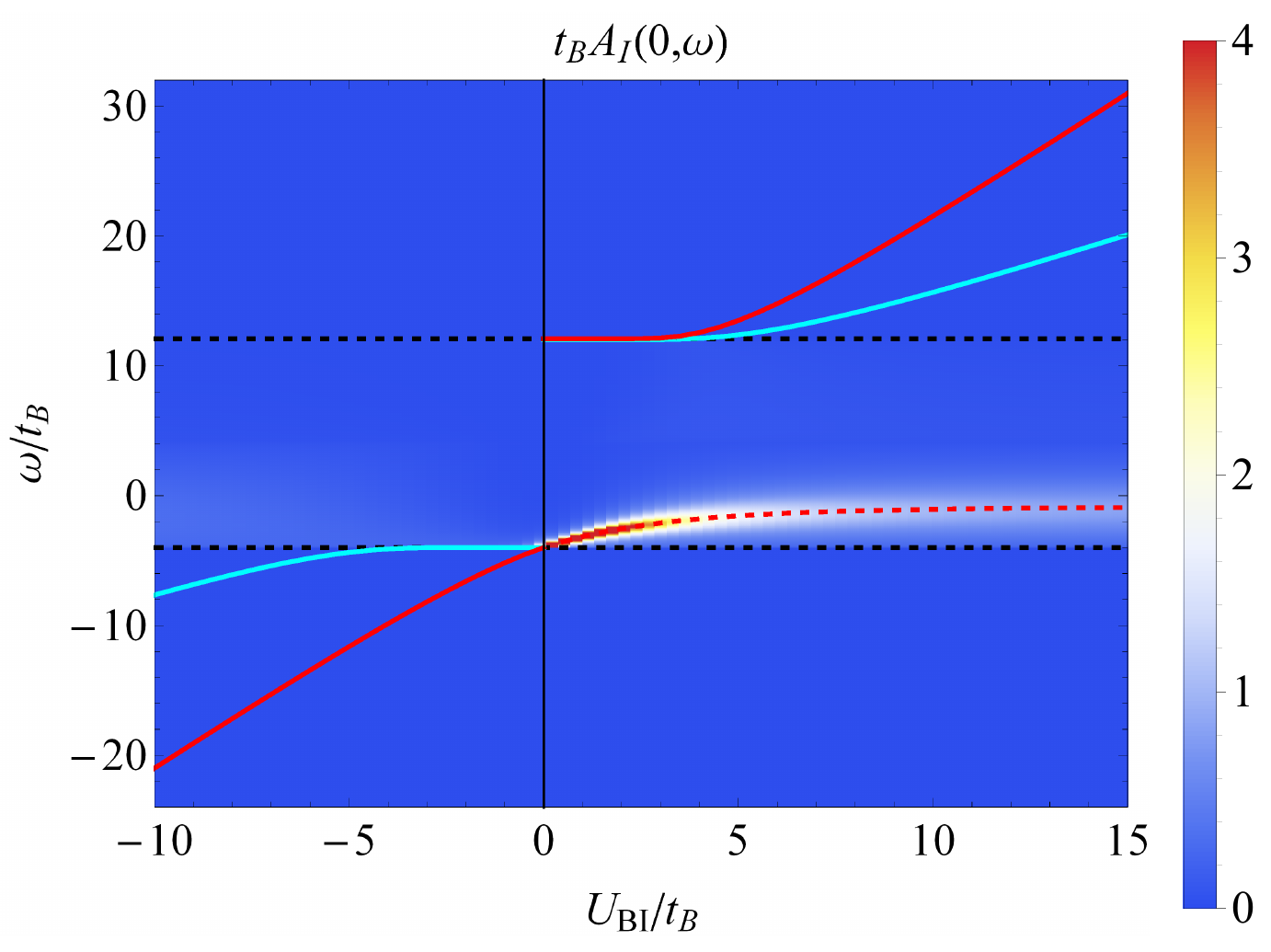}
\includegraphics[width=0.7\textwidth]{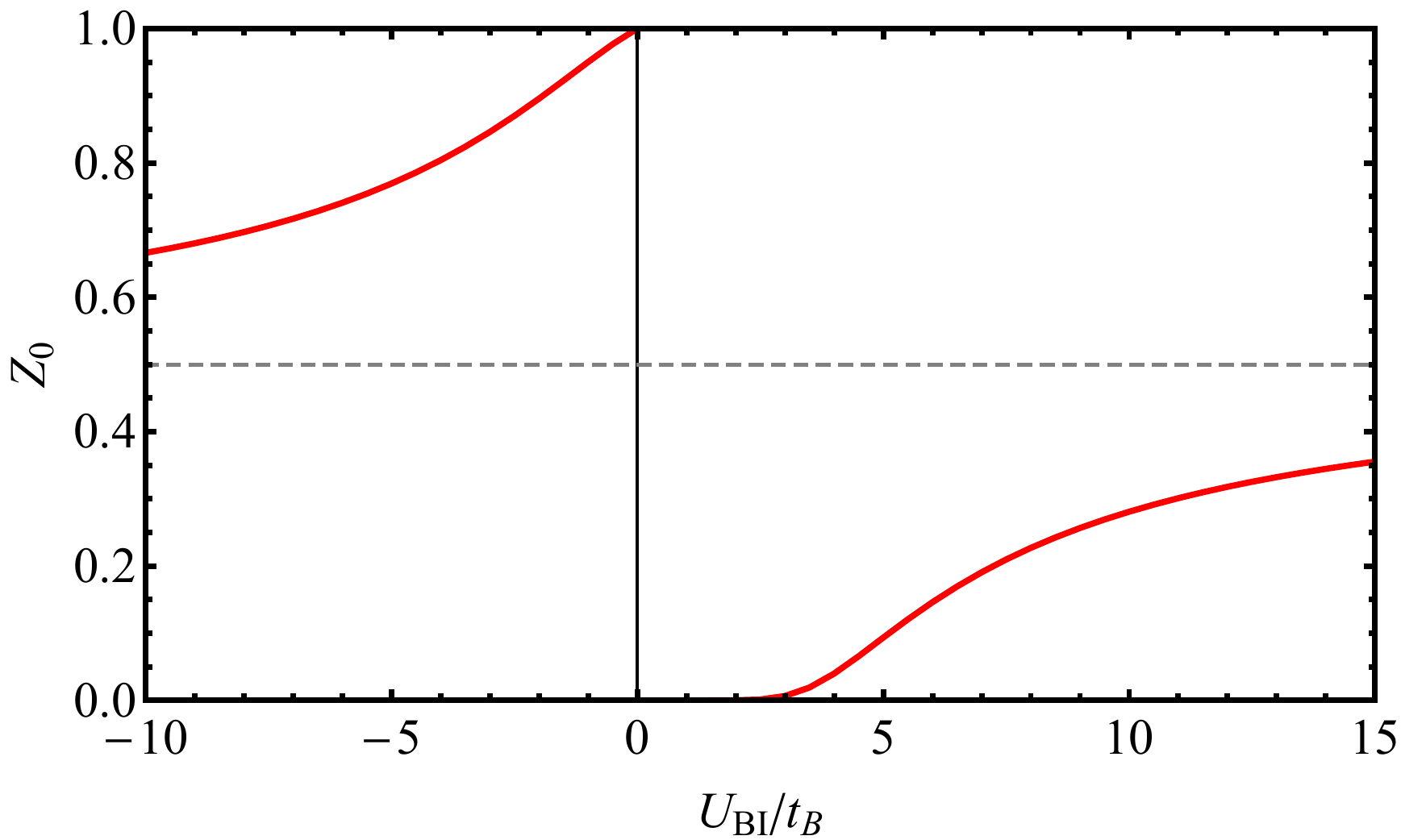}
\caption{The top panel shows the bound state energies (cyan lines) of one impurity  and one boson in an otherwise empty lattice as a function of $U_{BI}$. It also
shows the spectral function of the impurity with the red solid lines indicating the energy of an undamped attractive ($U_{BI}<0$) and upper polaron ($U_{BI}>0$). The dashed
red line gives the energy of the damped repulsive polaron, which is inside the single particle-Bogoliubov mode continuum indicated by the horizontal black dashed lines. The
bottom panel shows the corresponding quasiparticle residues, where the horizontal gray dashed line shows the asymptotic value of the residue for
 $|U_{BI}|\rightarrow\infty$ given by Eq.~\eqref{PolaronLargeUIZ}.}
\label{FigS}
\end{figure}

 The top panel of Fig.~\ref{FigS} shows the energy $\epsilon_B$ of such bound states as a function of $U_{BI}$ (cyan lines).
 For an attractive impurity-boson interaction $U_{BI}<0$ there is a  bound state with energy below the single particle continuum that has
 energies between  the black
 dashed horizontal lines. This is the
lattice analogue of the bound dimer state for homogeneous atomic gases in the BEC-BCS cross-over~\cite{Giorgini2008}.
Figure~\ref{FigS} furthermore shows that there is  a second bound state for repulsive interaction $U_{BI}>0$ with an energy above the single particle continuum. This
bound  state is stable since it has no available decay channels as  the single particle continuum is bounded from above~\cite{Piil2007,Valiente_2008}.
Such repulsively bound states have been observed for bosons in an optical lattice~\cite{winkler2006repulsively}. As shown in App.~\ref{App-Sec-TB-Bound}, it follows from the
two-dimensionality of the lattice that there is  a
two-body bound state  for any non-zero value of $U_{BI}$ either above ($U_{BI}>0$) or below ($U_{BI}<0$) the single-particle continuum.
For large $\omega$, we have $\Pi_{\text{v}}(\mathbf 0,\omega)\simeq 1/8zt_B$ so that $\epsilon_B\rightarrow U_{BI}+4t_B$ for $|U_{BI}|\rightarrow\infty$.

\section{Polarons \label{Sec-singleP}}
We now analyse the properties of a single impurity in the BEC and the formation of quasiparticles.
First, we focus on the spectral properties of the impurity, and then we discuss the spatial correlations between the impurity and the bosons.

\subsection{Spectral properties}
The spectral properties of an impurity with crystal momentum $\mathbf{k}$ immersed in the BEC can be  described by the impurity Green's function
\begin{equation}
G_I\left(i\omega,\mathbf{k}\right)=\frac{1}{i\omega-\epsilon_{I\mathbf{k}}-\Sigma(i\omega,\mathbf{k})},
\end{equation}
where $\Sigma(i\omega,\mathbf{k})$ is its self-energy.
The energy $\omega_\mathbf{k}$ of a quasiparticle is given by the pole of $G_I(i\omega,\mathbf{k})$, that is
$\omega_{\mathbf{k}}-\epsilon_{I\mathbf{k}}-\mathrm{Re}\Sigma(\omega_{\mathbf{k}}+i\eta,\mathbf{k})=0$
($\eta$ is a positive infinitesimal number) and the corresponding quasiparticle residue is $Z_{\mathbf{k}}=1/[1-\partial_{\omega}\mathrm{Re}\Sigma(\omega+i\eta,\mathbf{k})]\big|_{\omega=\omega_{\mathbf{k}}}$.

We use the ladder approximation to calculate the impurity self-energy~\cite{Rath2013}, which  has turned out to be in
 good agreement with experiments both for  equilibrium~\cite{Jorgensen2016,Ardila2019}
and non-equilibrium~\cite{Skou:2021aa} properties of the Bose polaron in continuum systems. In this approximation, the self-energy is
\begin{equation}
\Sigma\left(i\omega,\mathbf{k}\right)=n_0\mathcal{T}\left(i\omega,\mathbf{k}\right)
\label{SigmaP}
\end{equation}
where the scattering matrix is given by Eq.~\eqref{ScatteringMatrix} with the pair propagator generalised from Eq.~\eqref{PairpropVac} to
take into account the presence of the BEC. This yields
  \begin{align}
  \Pi(\mathbf P, i\Omega)=\frac{1}{M}\sum_{\mathbf{k}}\frac{u_{\mathbf k}^2}{i\Omega-\epsilon_{I\mathbf{P}-\mathbf{k}}-E_{\mathbf{k}}}
\label{Tmatrix}
\end{align}
at zero temperature,
where $u_{\mathbf{k}}^2=[(\epsilon_{B\mathbf{k}}+n_0U_B)/E_{\mathbf{k}}+1]/2$. Details of deriving Eq.~\eqref{Tmatrix} are given in
 App.~\ref{App-Sec-Pi-BEC}.

In the upper panel of Fig.~\ref{FigS}, we plot the spectral function $A_I(\mathbf k,\omega) =-2\text{Im} G_I(\mathbf k,\omega+i\eta)$ of the impurity for vanishing crystal momentum
 $\mathbf k=0$ as a function of $U_{BI}$. Here and in the rest of the paper we take the value $n_0=1$ and $U_B/t_B=0.07$ for the numerical calculations.
This value is experimentally realistic and ensures that the system is far away from the transition to the Mott phase~\cite{Spielman2008, Spielman2010}.
 Figure \ref{FigS} shows that  there is a well-defined
 quasiparticle branch below  the dimer state energy for $U_{BI}<0$. Its quasiparticle residue decreases from unity with
 decreasing $U_{BI}$ as   shown in the lower panel of Fig.~\ref{FigS}.
 The  energy of this quasiparticle branch smoothly increases to positive values when  $U_{BI}>0$,  where it is  broadened
due to decay into the single particle-Bogoliubov mode continuum with energies $\epsilon_{I\mathbf k}+E_{-\mathbf k}$, which is almost indistinguishable from the single particle continuum in Fig.~\ref{FigS}. This quasiparticle branch is the lattice analogue of the attractive polaron in continuum gases that smoothly
evolves into a damped repulsive polaron for positive impurity-boson interaction.  We note that the
  broadening shown in Fig.~\ref{FigS} may be an artefact of the approximation used. Indeed,  a self-consistent approach would move the single particle-Bogoliubov mode continuum
  above the energy of the  repulsive polaron for ${\mathbf k}=0$~\cite{Massignan:2011aa}.

Figure~\ref{FigS} furthermore shows a  second quasiparticle branch  above the single particle-Bogoliubov mode continuum for repulsive impurity-boson interaction.
 It has  an  energy \emph{above} that of the repulsive dimer and a  residue increasing from zero with $U_{BI}$,
 as can be seen in the lower panel of Fig.~\ref{FigS}. We denote this quasiparticle branch with infinite lifetime, which has no analogue for continuum gases,
  as the {\em upper polaron}.
 For large frequencies, we can approximate the pair propagator by  $\Pi(\omega)=A/\omega$ with $A=\sum_{\mathbf{k}}u_{\mathbf k}^2/M$  a coefficient determined by the medium.
 This in turn gives the residue and energy
\begin{subequations}
\begin{equation}
 Z_{\mathbf k}\rightarrow\frac{n_0}{n_0+A}
\label{PolaronLargeUIZ}
\end{equation}
\begin{equation}
\omega_{\mathbf k}\rightarrow(n_0+A)U_{BI}
\label{PolaronLargeUIomega}
\end{equation}
\label{PolaronLargeUI}
\end{subequations}
of the attractive and upper polaron  for  $U_{BI}\rightarrow-\infty$ and $U_{BI}\rightarrow\infty$ respectively.
Equation \eqref{PolaronLargeUIZ} shows that for  $|U_{BI}|\rightarrow\infty$ the residue of the polaron saturates  to a non-zero value
 that can be tuned by  changing the density of the BEC.  This value is shown as a horizontal dashed line in
 Fig.~\ref{FigS} (bottom). For an ideal BEC with density $n_0=1$, Eq.~\eqref{PolaronLargeUIZ} predicts $ Z_{\mathbf k}\rightarrow 1/2$ whereas we
 obtain  $ Z_{\mathbf k}\rightarrow 0.4993$ for $U_B/t_B=0.07$.

\begin{figure}[h]
\centering
\includegraphics[width=0.98\textwidth]{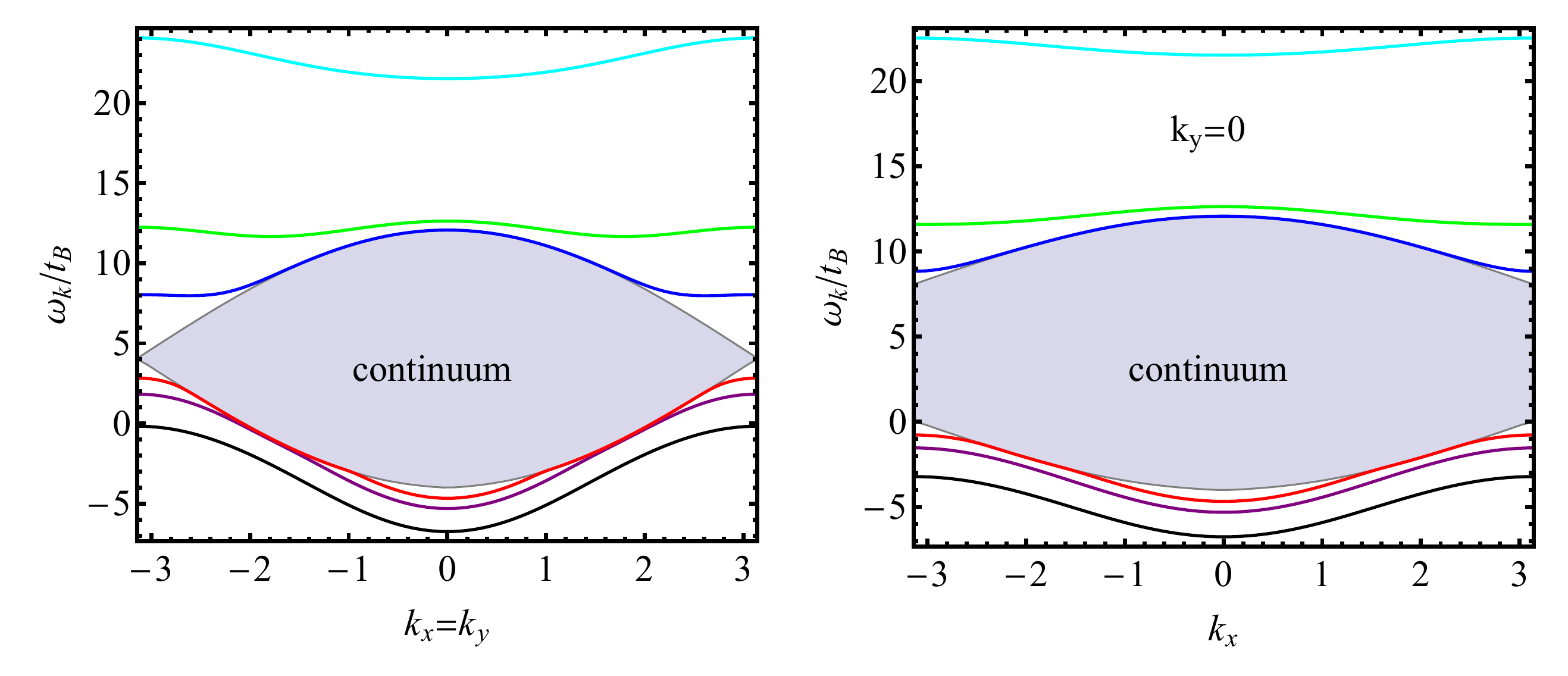}%
\caption{The energies of the attractive polaron for $U_{BI}<0$  and upper polaron for  $U_{BI}>0$ as a function of the crystal
momentum along the diagonal $k_x=k_y$ (left panel) and along the $k_x$-axis with $k_y=0$ (right panel). The lines from bottom to top are for $U_{BI}/t_B=-2.1$ (black),  $U_{BI}/t_B=-1.1$ (purple),
 $U_{BI}/t_B=-0.6$ (red),  $U_{BI}/t_B=2$ (blue),  $U_{BI}/t_B=4.1$ (green), and  $U_{BI}/t_B=10$ (cyan). The single particle-Bogoliubov mode continuum is
 indicated by the gray region.
}
\label{Figwkdiag}
\end{figure}
In Fig.~\ref{Figwkdiag}, we plot the polaron energy $\omega_\mathbf{k}$   as a function of the crystal momentum along the diagonal $k_x=k_y$ (left panel) and along the $k_x$-axis with $k_y=0$ (right panel) in the BZ for
 different values of the impurity-boson interaction $U_{BI}$. The single  particle-Bogoliubov mode continuum with energies
in the interval $\min_{\mathbf 	q}(\epsilon_{I\mathbf q}+E_{\mathbf k-\mathbf q})\le\omega\le\max_{\mathbf 	q}(\epsilon_{I\mathbf q}+E_{\mathbf k-\mathbf q})$ is also
shown. When $U_{BI}<0$, we plot the energy of
 the attractive polaron, and when $U_{BI}>0$ we plot the energy of the upper polaron. Figure \ref{Figwkdiag} shows that
 the energy band of the attractive polaron below the continuum has a minimum at ${\mathbf k}=0$, whereas the energy band of the upper polaron has a more
 complicated structure with a local maximum ${\mathbf k}=0$ that evolves into a minimum for increasing impurity-boson repulsion.

\begin{figure}[h]
\centering
\includegraphics[width=0.9\textwidth]{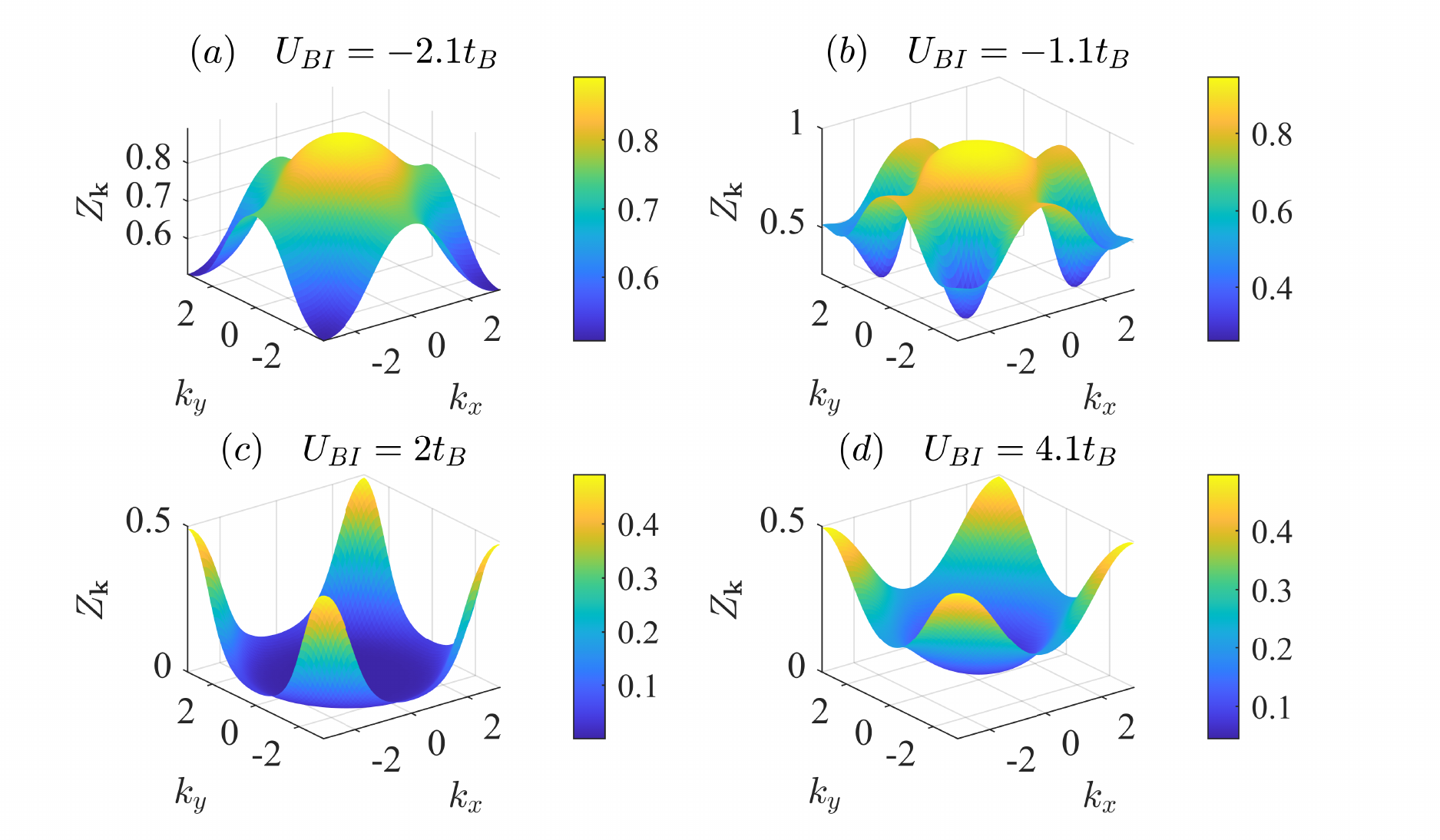}%
\caption{The quasiparticle residue  of the attractive polaron  for $U_{BI}/t_B=-2.1$ and $U_{BI}/t_B=-1.1$, and for the upper polaron
for $U_{BI}/t_B=2$ and $U_{BI}/t_B=4.1$ in the BZ.}
\label{Fig-Zk}
\end{figure}

Figure \ref{Fig-Zk} shows the quasiparticle residue of the  polaron  in the BZ for
different values of $U_{BI}$. When $U_{BI}/t_B=-2.1$ shown in panel (a), the residue has a maximum at ${\mathbf k}=0$ and minima at the corners of the BZ,
and it is larger than $1/2$ for all $\mathbf k$. For the smaller attractive interaction $U_{BI}/t_B=-1.1$ shown  in panel (b),
the minima have moved to $k_x=k_y\simeq2$ and correspond to a small residue, reflecting that the energy of the attractive polaron is close to the single particle-Bogoliubov mode continuum.
Contrary to this, the residue of the upper polaron has maxima at the corners of the BZ whereas it is very small close to zero momentum
as can be seen in panel (c) for $U_{BI}/t_B=2$. By comparing
with panel (d), we see that the  residue of the upper polaron  increases and is  non-zero in the whole BZ for larger impurity-boson repulsion.


 \subsection{Spatial impurity-boson correlations}
 Inspired by the impressive single site resolution microscopy available in optical lattices~\cite{Bakr:2009aa,Sherson:2010aa},
 we now turn to the spatial correlations between the impurity and the bosons in the polaron states.

  The spatial properties are most straightforwardly calculated from the polaron wave function, which
in the ladder approximation reads
\begin{gather}
|\Psi_P\rangle=(\phi_0\hat c^\dagger_{\mathbf{k}=0}+\sum_{\mathbf k}\psi_{\mathbf k}\hat c^\dagger_{\mathbf k}\hat \beta^\dagger_{-\mathbf k})|\text{BEC}\rangle.
\label{Chevy}
\end{gather}
Here $\hat c^\dagger_{\mathbf{k}}/\hat \beta^\dagger_{\mathbf k}$ creates an impurity/Bogoliubov mode with crystal momentum ${\mathbf k}$,
 $|\text{BEC}\rangle$ denotes the ground state of the BEC defined by $\hat \beta_{\mathbf k}|\text{BEC}\rangle=0$ and we have taken the momentum
of the polaron to be zero.
The variational parameters  $\phi_0$ and $\psi_{\mathbf k}$ are determined by minimising the energy $\langle\Psi_P|\hat H|\Psi_P\rangle$, and one can show that they
are directly obtained from the ladder calculation as  $\phi_0=\sqrt{Z_{0}}$ and {$\psi_{\mathbf k}=\sqrt{Z_0}u_{\mathbf k}\left(\omega_{\mathbf 0}-\epsilon_{I\mathbf 0}\right)/\sqrt{nM}(\omega_{\mathbf{0}}-\epsilon_{I\mathbf k}-E_{\mathbf k})$},  see App.~\ref{SinglePolaronAppendix}.

To analyse the spatial correlations between the impurity and the surrounding BEC, we calculate
\begin{equation}
\mathcal C({\mathbf i}-{\mathbf j})=M\langle \Psi_P| [\hat n_B({\mathbf i})-n]\hat n_I({\mathbf j})|\Psi_P\rangle
\label{CorrFn}
\end{equation}
where $\hat n_B({\mathbf i})=\hat b_{\mathbf i}^\dagger\hat b_{\mathbf i}$ and $\hat n_I({\mathbf i})=\hat c_{\mathbf i}^\dagger\hat c_{\mathbf i}$,
$n=\langle \Psi_P| \hat n_B({\mathbf i})|\Psi_P\rangle$ gives the total number of bosons at lattice site ${\mathbf i}$,  and we have assumed an infinite lattice.
 The correlation function $\mathcal C({\mathbf i}-{\mathbf j})$ gives the number of bosons at site ${\mathbf i}$ given that the impurity is at site ${\mathbf j}$.
 As detailed in App.~\ref{SinglePolaronAppendix}, a straightforward calculation using Bogoliubov theory yields
\begin{equation}
\label{Cr}
\mathcal C(\mathbf i)=\sqrt{\frac{n}{M}}\sum_{\mathbf k}\left[\phi_0^*\psi_{\mathbf k}(u_{\mathbf k}+v_{\mathbf k}) e^{i\mathbf k\cdot\mathbf i}+\text{h.c}\right]
+\frac{1}{M}\sum_{\mathbf k,\mathbf k'}\psi_{\mathbf k}^*\psi_{\mathbf k'}
(u_{\mathbf k}u_{\mathbf k'}+v_{\mathbf k}v_{\mathbf k'})e^{i(\mathbf k-\mathbf k')\cdot\mathbf i}.
\end{equation}

\begin{figure}[ht]
\centering
\includegraphics[width=0.7\textwidth]{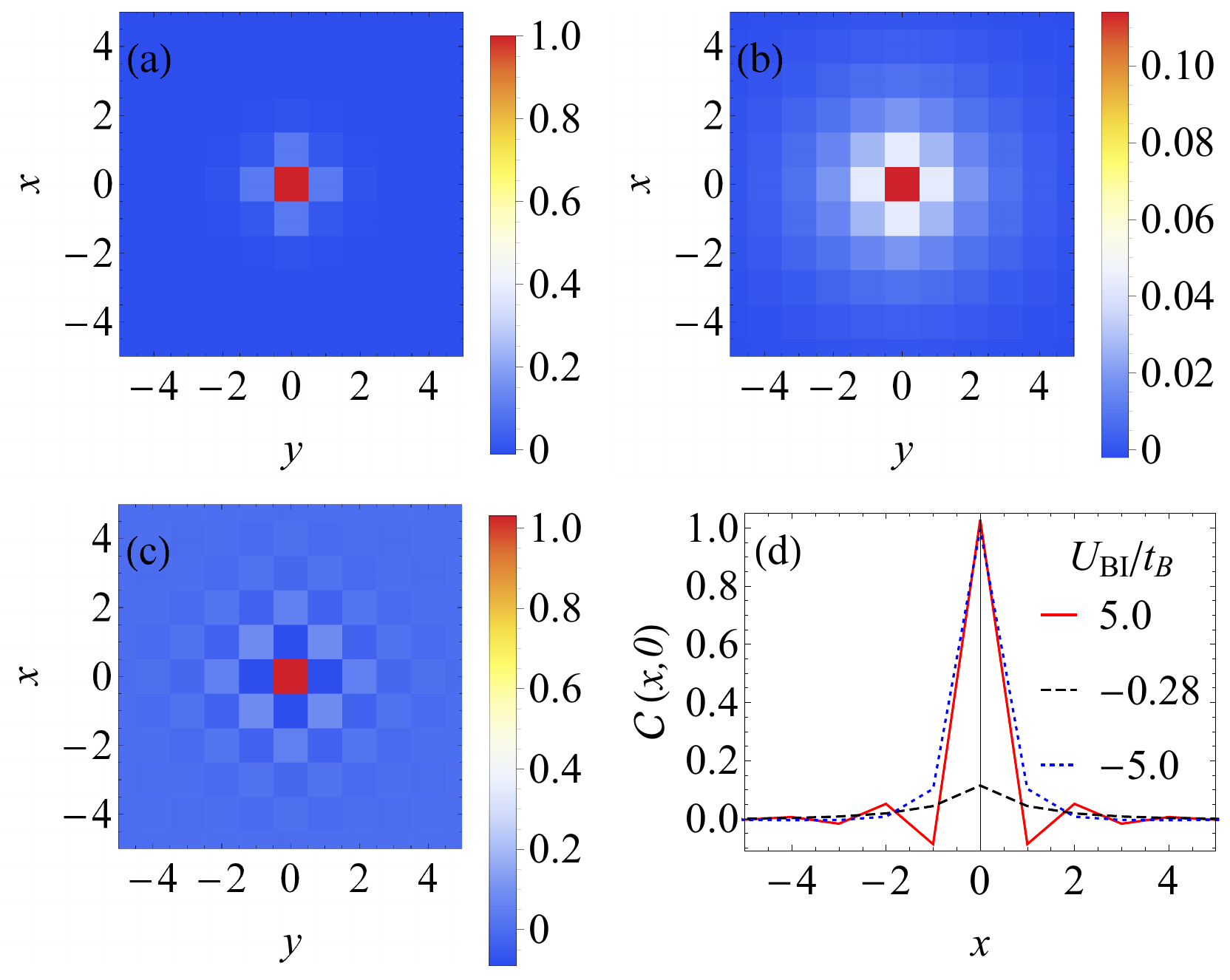}%
\caption{
The spatial correlations between the impurity and the bosons as described by $\mathcal C(\mathbf i)$ in the attractive polaron state
for $U_{BI}/t_B=-5$ (a) and  $U_{BI}/t_B=-0.28$ (b), and in the upper polaron state for
$U_{BI}/t_B=5.0$ (c). Panel (d) shows $\mathcal C(x,0)$ for the three impurity-boson interaction strengths.}
\label{SpatialC}
\end{figure}
Figure \ref{SpatialC} shows the spatial correlation function $\mathcal C(\mathbf i)$ for
  different values of $U_{BI}$.  First, for a strong and attractive impurity-boson interaction $U_{BI}/t_B=-5$ shown in panel (a), the number of bosons
 at the lattice site of the impurity is increased by one in the attractive polaron state,
 whereas it quickly relaxes to the background value away from the impurity. This corresponds to a
  strongly localised dressing cloud  containing approximately one boson on top of the impurity. As expected, the number of bosons in the
 dressing cloud  decreases and it becomes more spatially extended with decreasing attraction as can be seen in panel (b) plotting
   $\mathcal C(\mathbf i)$ for $U_{BI}/t_B=-0.28$.
 For a strong repulsive interaction $U_{BI}/t_B=5$ shown in panel (c), the dressing cloud is again localised at the impurity and contains approximately
 one boson. This is because it is energetically forbidden for the boson to tunnel to a neighbouring site in the upper polaron state.
  Finally, we plot in panel (d)  $\mathcal C(x,0)$  for the three
  different bose-impurity interaction strengths, illustrating further how the dressing cloud becomes more
 localised both for large attractive and repulsive impurity-boson interaction.  In addition, we see that $\mathcal C(x,0)$ is negative at the sites neighbouring the impurity
  when
  $U_{BI}/t_B=5$. Physically, this means that while the impurity attracts a boson to its own lattice site in the upper polaron state, it pushes them away from its nearest
  neighbour sites contrary to the case of the attractive polaron where $\mathcal C(\mathbf i)$ is positive everywhere.

\section{Mediated interaction and effective Schr\"odinger equation}
Having analysed individual polarons, we are now ready to explore how they interact  by the exchange of Bogoliubov modes, and to
 derive an effective Schr\"odinger equation for two polarons in the presence of the BEC.

\subsection{Schr\"odinger equation}
It is convenient to derive an effective Schr\"odinger equation to describe the effects of a general interaction between two polarons. To do  this,
we consider the scattering of two impurities with momenta and energies $p_1$ and $p_2$
to momenta and energies $p_1+q$ and $p_2-q$. The corresponding scattering matrix obeys the Bethe-Salpeter equation, which in the ladder approximation
reads
\begin{align}
\Gamma(p_1,p_2;q)=V(p_1,p_2;q)-\frac{T}{M}\sum_{q_1}V(p_1,p_2;q_1)\times\nonumber\\
G_I(p_1+q_1)G_I(p_2-q_1)\Gamma(p_1+q_1,p_2-q_1;q-q_1).
\label{Eq-BS}
\end{align}
Here, $V(p_1,p_2;q_1)$ is the interaction between the impurities, $T$ the temperature, and the sum  is over momenta in the BZ and Matsubara
frequencies. Since an interaction mediated by Bogoliubov modes
 is not Galilean invariant and the speed of sound in the BEC is finite, $V(p_1,p_2;q_1)$ in
general depends on all momenta and energies. In order to simplify the problem,  we therefore make some approximations.
First, we use a pole expansion for the impurity Green's function around the polaron energies $\omega_{\mathbf k}$ writing
$G_I(i\omega,\mathbf{k})\simeq Z_\mathbf{k}/(i\omega-\omega_{\mathbf{k}})$ and multiply both sides of Eq.~\eqref{Eq-BS} by $Z_{\mathbf{p}_1} Z_{\mathbf{p}_2}$.
This  approximation corresponds to going to a quasiparticle picture with
 $\Gamma_{\mathrm{eff}}(p_1,p_2;q)=Z_{\mathbf{p}_1}Z_{\mathbf{p}_2}\Gamma(p_1,p_2;q)$ the scattering matrix between
 two polarons with a   quasiparticle interaction $V_{\mathrm{qp}}(p_1,p_2;q)=Z_{\mathbf{p}_1}Z_{\mathbf{p}_2}V(p_1,p_2;q)$~\cite{orland_quantum_1998}.
 Second,  we neglect the dependence of the induced interaction on the frequency transfer, which is accurate
  when the binding energy of the bipolaron is small compared to the typical excitation energy of the Bogoliubov modes. Finally, we take the energy of the
  scattering particles to be a constant determined by the energy of the incoming polarons.   Using these approximations,
which are reliable as long as there are well-defined   polarons with a large spectral weight and the bi-polarons are weakly bound,
  the frequency sum in Eq.~\eqref{Eq-BS} can now be evaluated analytically. The result is a Lippmann-Schwinger equation that in turn
 is equivalent to an effective Schr\"odinger equation for  two polarons in the BEC as explained in App.~\ref{App-Sec-Vind}.
 Taking the COM momentum of the two polarons to be zero,  the Schr\"odinger equation reads
\begin{equation}
(E_B-2\omega_{\mathbf{k}})\psi(\mathbf{k})=\frac{1}{M}\sum_{\mathbf{k}'}V_{\mathrm{qp}}(\mathbf{k},\mathbf{k}')\psi(\mathbf{k}').
\label{Eq-eff-SE}
\end{equation}
Here, $E_B$ is the energy
of the two-polaron state with  wave function $\psi(\mathbf{k})$, where $\mathbf{k}$ is
their relative momentum.
We have changed notation so that $V_{\mathrm{qp}}(\mathbf{k},\mathbf{k}')$ denotes the interaction between two polarons with zero COM momentum scattering from relative momentum
${\mathbf k}$ to ${\mathbf k}'$. Since the interaction depends on both the in-coming and out-going relative momentum, it
 is non-local  reading   $\sum_{\mathbf i'}V({\mathbf i},{\mathbf i}')\psi({\mathbf i}')/M$ in real space. This
  is typical of effective two-body Schr\"odinger equations in many-body systems such as for instance the Skyrme force in nuclear matter~\cite{ring2004nuclear}.

\subsection{Polaron-polaron interaction}
There are two contributions to the quasiparticle interaction between two polarons, which we write as
\begin{equation}
V_{\mathrm{qp}}(\mathbf{k},\mathbf{k}')=Z_{\mathbf{k}}^2U_I+V_{\mathrm{ind}}(\mathbf{k},\mathbf{k}').
\label{QPint}
\end{equation}
The first term comes from the direct interaction between the  impurities, which is reduced by a factor $Z_{\mathbf{k}}^2$ since this is
 the impurity component of the two polarons. The second term is the induced interaction between the polarons
mediated by Bogoliubov modes in the BEC. As shown in App.~\ref{App-Sec-Vind},
this interaction is given by
\begin{gather}
V_{\mathrm{ind}}(\mathbf{k},\mathbf{k}')=Z_{\mathbf{k}}^2n_0[2\mathcal{T}(\mathbf{k},\omega)G_{11}(\mathbf{k}'-\mathbf{k},0)\mathcal{T}(\mathbf{k}',\omega)+
\nonumber\\
\mathcal{T}^2(\mathbf{k}',\omega)G_{12}(\mathbf{k}'-\mathbf{k},0)+\mathcal{T}^2(\mathbf{k},\omega)G_{12}(\mathbf{k}'-\mathbf{k},0)]
\label{InducedInt}
\end{gather}
 to leading order in the number of Bogoliubov  modes.  Here,
 $G_{ij}(\mathbf{k},0)$ are the  Green's functions of the BEC, and $\omega$ is taken to be the energy of the interacting quasiparticles.
 The scattering matrices in Eq.~\eqref{InducedInt} take the two-body  impurity-boson correlations into account exactly.

\begin{figure}[h]
\centering
\includegraphics[width=0.7\textwidth]{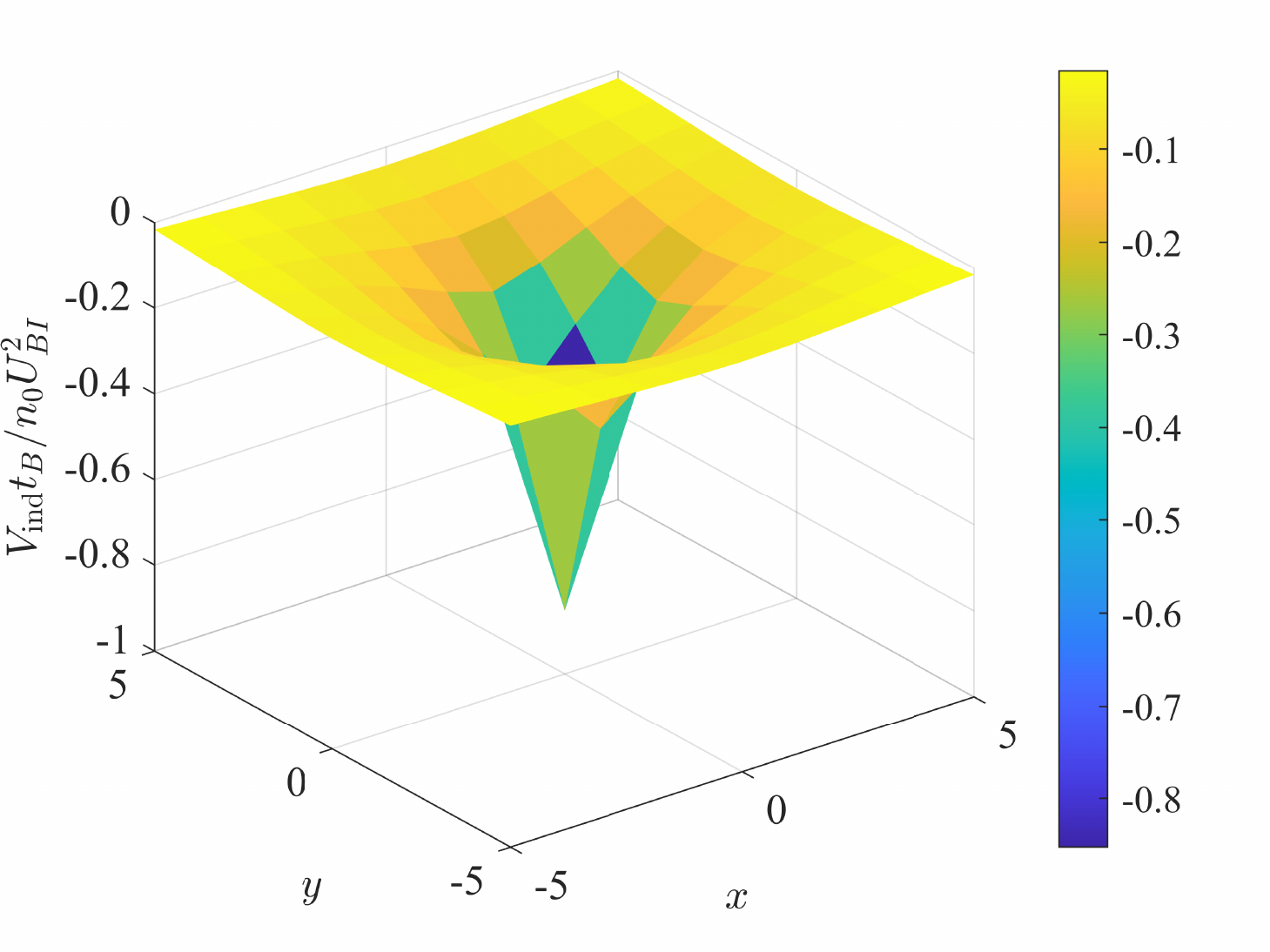}%
\caption{The induced interaction $V_{\mathrm{ind}}(\mathbf{i})$ between two polarons in real space obtained by Fourier transforming Eq.~\eqref{InducedPert}.}
\label{Fig:InducedInt}
\end{figure}

For weak impurity-boson interaction, we have $\mathcal{T}\simeq U_{BI}$ and the
induced interaction simplifies to
\begin{equation}
V_{\mathrm{ind}}(\mathbf{k},\mathbf{k}')=-\frac{2n_0U_{BI}^2\epsilon_{B\mathbf{k}-\mathbf{k}'}}{E^2_{\mathbf k-\mathbf{k}'}}.
\label{InducedPert}
\end{equation}
This shows that it  to leading order is proportional to $U_{BI}^2$. In Fig.~\ref{Fig:InducedInt}, we plot the induced interaction in real space
 obtained by Fourier transforming Eq.~\eqref{InducedPert}.
  This illustrates the attractive nature of the induced interaction and that its range is a few lattice sites.
Approximating the single particle  dispersion at small momenta as quadratic and Fourier transforming Eq.~\eqref{InducedPert} to real space yields
  the long range behaviour $V_{\mathrm{ind}}(r)\propto\exp(-\sqrt{2n_0U_B/t_B}r)/\sqrt{r}$.

While we have employed a number of approximations in deriving the  Schr\"odinger equation   Eq.~\eqref{Eq-eff-SE} with the quasiparticle
interaction given by Eqs.\eqref{QPint}-\eqref{InducedInt}, we note that a similar approach
turns out to be  remarkably accurate  when compared with Monte-Carlo calculations for continuum quantum gases,
 even for strong impurity-boson interactions leading to large bipolaron binding energies~\cite{camacho2018bipolarons}.

\section{Bipolarons}
We can now address  question whether the induced interaction  is
strong enough to lead to  bound states between two polarons. Note that while the condition for two-body bound states in a 2D continuum system  with a local interaction $V(r)$ is
$\int\!drrV(r)<0$~\cite{landau1981quantum}, the present case is more complicated since we are dealing with a non-local effective interaction coming from
integrating out the bosonic degrees of freedom in a lattice BEC.
We focus on bound states between two  attractive polarons  taking $U_{BI}<0$, and we therefore use the attractive polaron energies
$\omega_{{\mathbf k}}$ in Eq.~\eqref{Eq-eff-SE}. Likewise, the quasiparticle interaction
is calculated from Eq.~\eqref{InducedInt} using the residues $Z_{\mathbf k}$ of the attractive polarons and the ground state energy $\omega=\omega_{\mathbf k=0}$.

\begin{figure}[!h]
\centering
\includegraphics[width=0.6\textwidth]{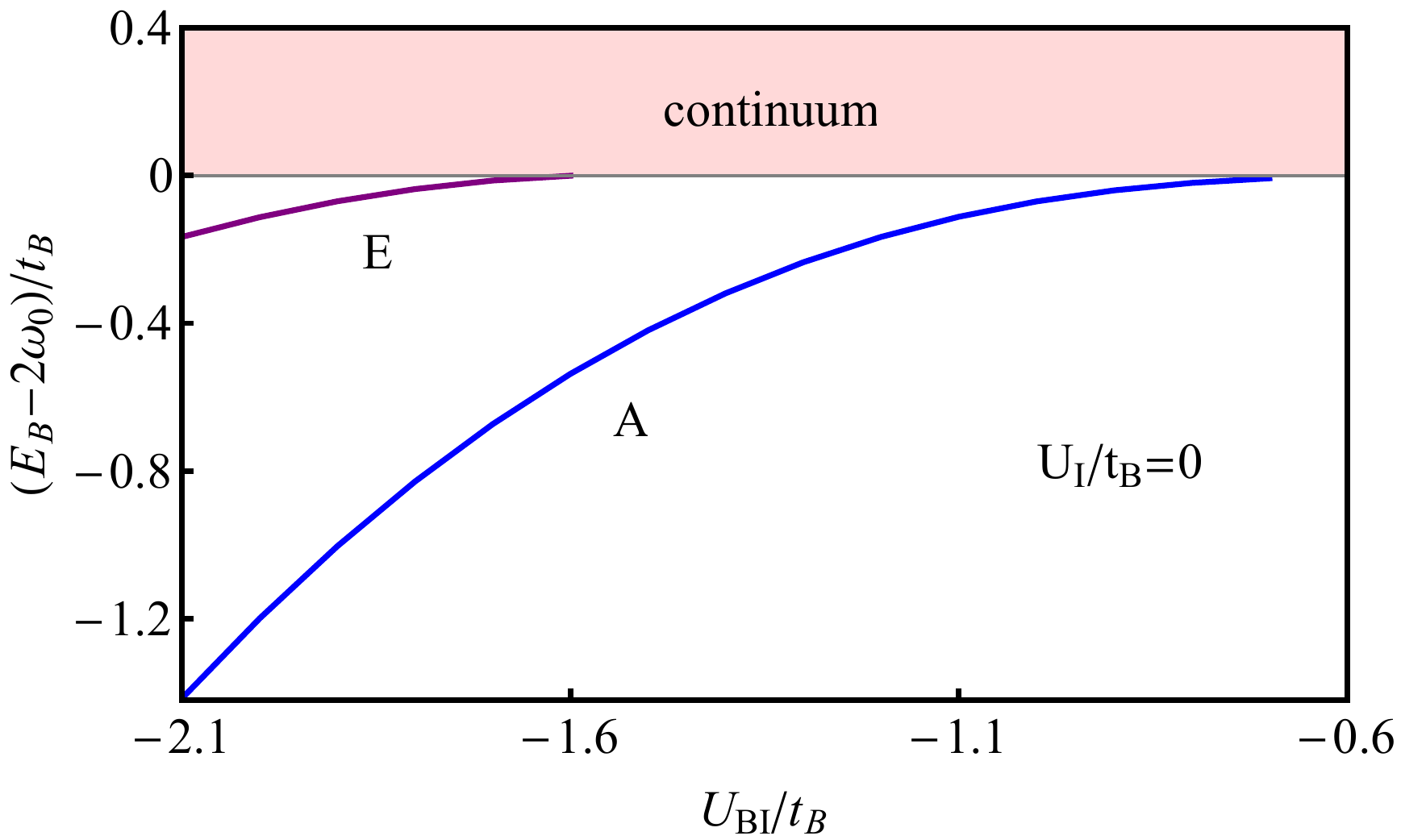}\\%
\includegraphics[width=0.6\textwidth]{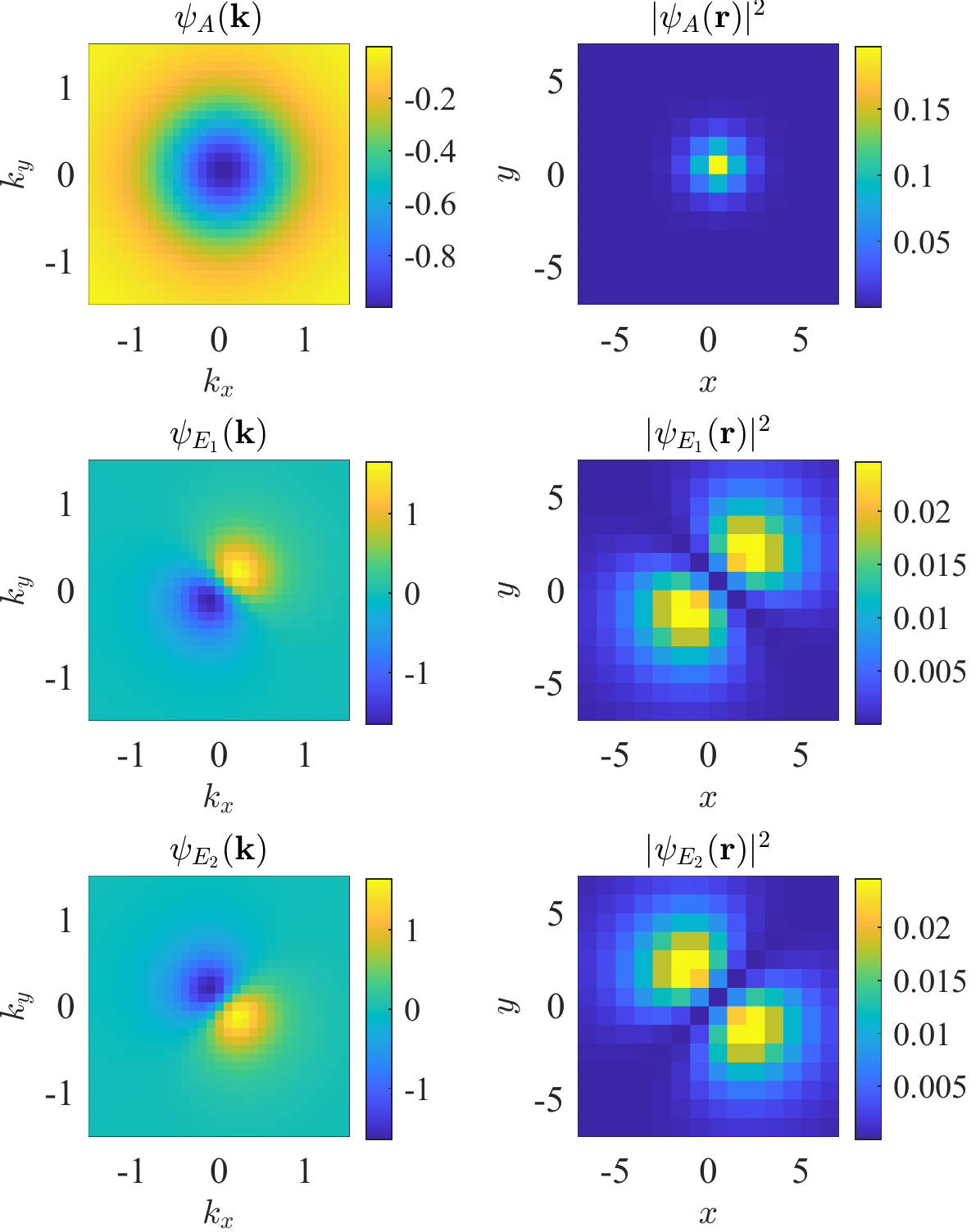}%
\caption{The top panel shows the energy of the bipolaron states relative to the  continuum of two unbound attractive polarons with zero COM momentum
as a function of $U_{BI}$ for $U_I=0$.  The blue line (A) is the energy of
the $C_4$ symmetric bipolaron relevant for bosonic impurities whereas the  purple line (E) gives the energy of the  two bipolaron states with $C_2$ symmetry relevant for fermionic
impurities. The bottom panel  shows the corresponding wave functions in momentum space and densities in real space for $U_{BI}/t_B=-2.1$.}
\label{Fig:EnergiesAttractive}
\end{figure}

In the top panel of Fig.~\ref{Fig:EnergiesAttractive}, we plot the eigenvalue spectrum obtained from solving Eq.~\eqref{Eq-eff-SE}  as a function of $U_{BI}$.
The bare impurities are taken to be non-interacting with $U_I=0$ so that
any bound states arise exclusively from the induced interaction. The solutions of Eq.~\eqref{Eq-eff-SE} fall in two classes. First, there is a set of scattering states
 with energies close to those of two free attractive polarons  $\omega_{\mathbf k}+\omega_{-\mathbf k}$,
which form a continuum in the thermodynamic limit. Since we have subtracted the ground state energy $2\omega_{\mathbf k=0}$
of two unbound attractive polarons, this continuum starts at zero energy in Fig.~\ref{Fig:EnergiesAttractive}. Second,
we see that for sufficiently large $|U_{BI}|$ a solution with a discrete energy separated from this continuum emerges. This solution corresponds to a
 bound dimer state, i.e.\ a bipolaron, with a binding energy increasing  with $|U_{BI}|$.
 The lower panel of Fig.~\ref{Fig:EnergiesAttractive} plots the corresponding wave function  $\psi_A({\mathbf k})$ in momentum space for $U_{BI}/t_B=-2.1$.
 It is centered in the BZ and is $C_4$ (``$s$-wave'') symmetric
due to the square lattice. We denote this bipolaron with the label ``A'' corresponding to the irreducible representation of the lattice symmetry
group $C_{4v}$ it spans~\cite{Atkins2008}. Since its wave function is symmetric under particle-exchange, i.e.\  a $C_2$ operation, it is
 relevant for bosonic impurities, whereas two fermionic impurities cannot occupy this bipolaron state.
Note that the fact that the wave function is centered at the origin in momentum space  makes the approximation using the
ground state polaron energy $\omega_{\mathbf k=0}$ in the mediated interaction consistent. In Fig.~\ref{Fig:EnergiesAttractive},  we also plot the
probability $|\psi_A({\mathbf r})|^2$ of finding the two impurities separated by the vector ${\mathbf r}$ in the lattice, which is
obtained by a Fourier transform of $\psi_A({\mathbf k})$.
The probability  is strongly localised, which means that the two impurities are mostly at the same or neighboring lattice sites when bound together in the bipolaron state.

Figure \ref{Fig:EnergiesAttractive}  shows that another negative energy branch emerges for  $U_{BI}/t_B\lesssim-1.6$. This branch in fact  corresponds to two
degenerate bipolaron states with wave functions $\psi_{E_1}$ and $\psi_{E_2}$
plotted in the lower panel of Fig.~\ref{Fig:EnergiesAttractive}  for  $U_{BI}/t_B=-2.1$. These two
degenerate states span the two-dimensional irreducible representation E of $C_{4v}$, which is the only one that is odd under
particle exchange. These two bipolaron states are therefore relevant for fermionic impurities in the BEC~\cite{Atkins2008}, and
their  wave functions have a ``$p$-wave'' symmetry with a node at the origin. As a result, they are considerably more extended in real space as compared
 to the lowest bipolaron state, meaning that the two impurities are most likely to be at neighbouring lattice sites whereas they are never at the same site.
 Note that one can form linear combinations of these two wave functions
 that are extended along the $x$- and $y$-axis instead of along the diagonals.

These results demonstrate that the mediated interaction between two polarons  is indeed strong enough to support bound states in the 2D lattice. The bipolarons
  should furthermore be  observable by looking at the spatial correlations between two impurities
  using the single site resolution of atomic microscopy in optical lattices.

\begin{figure}[!h]
\centering
\includegraphics[width=0.65\textwidth]{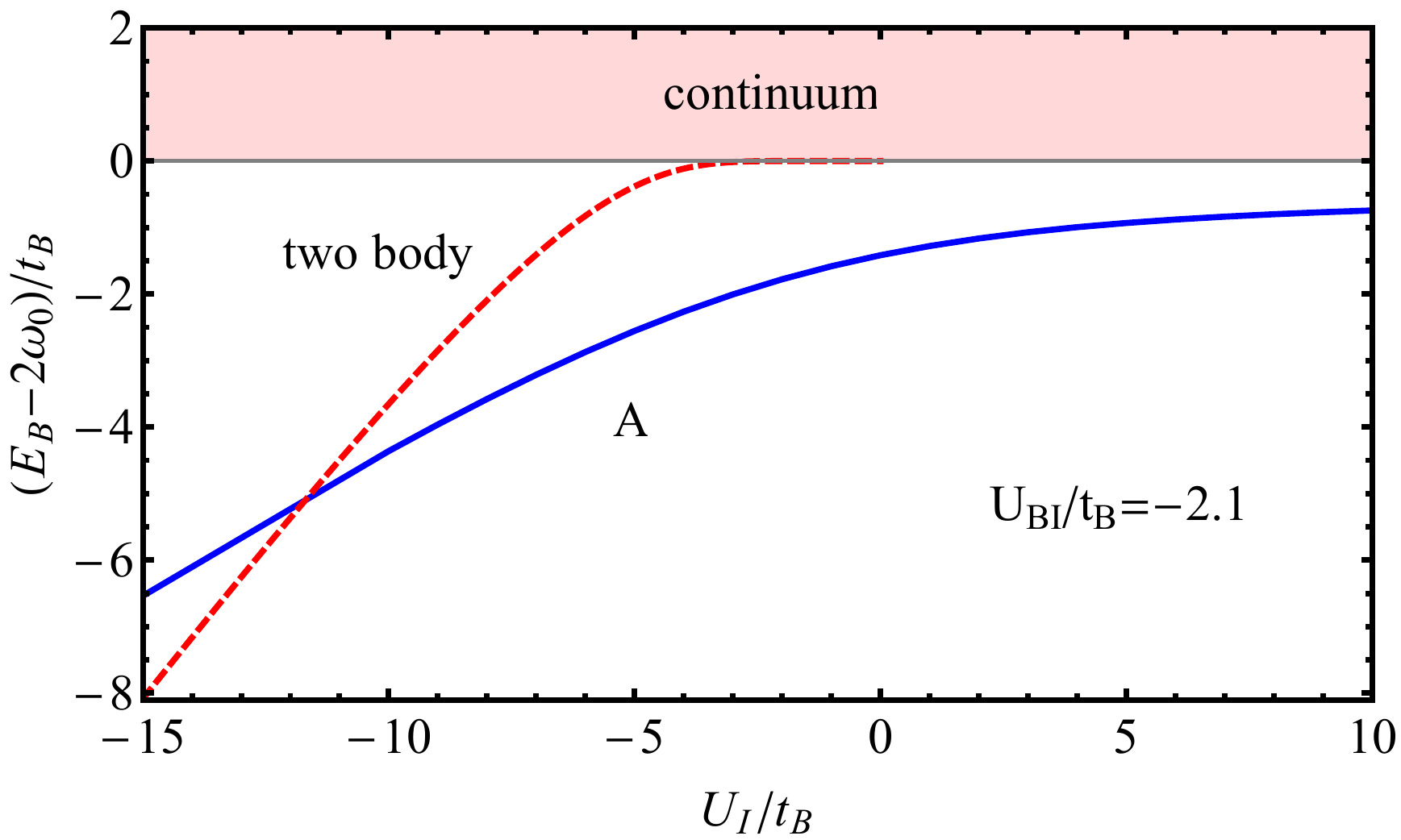}\\%
\includegraphics[width=0.65\textwidth]{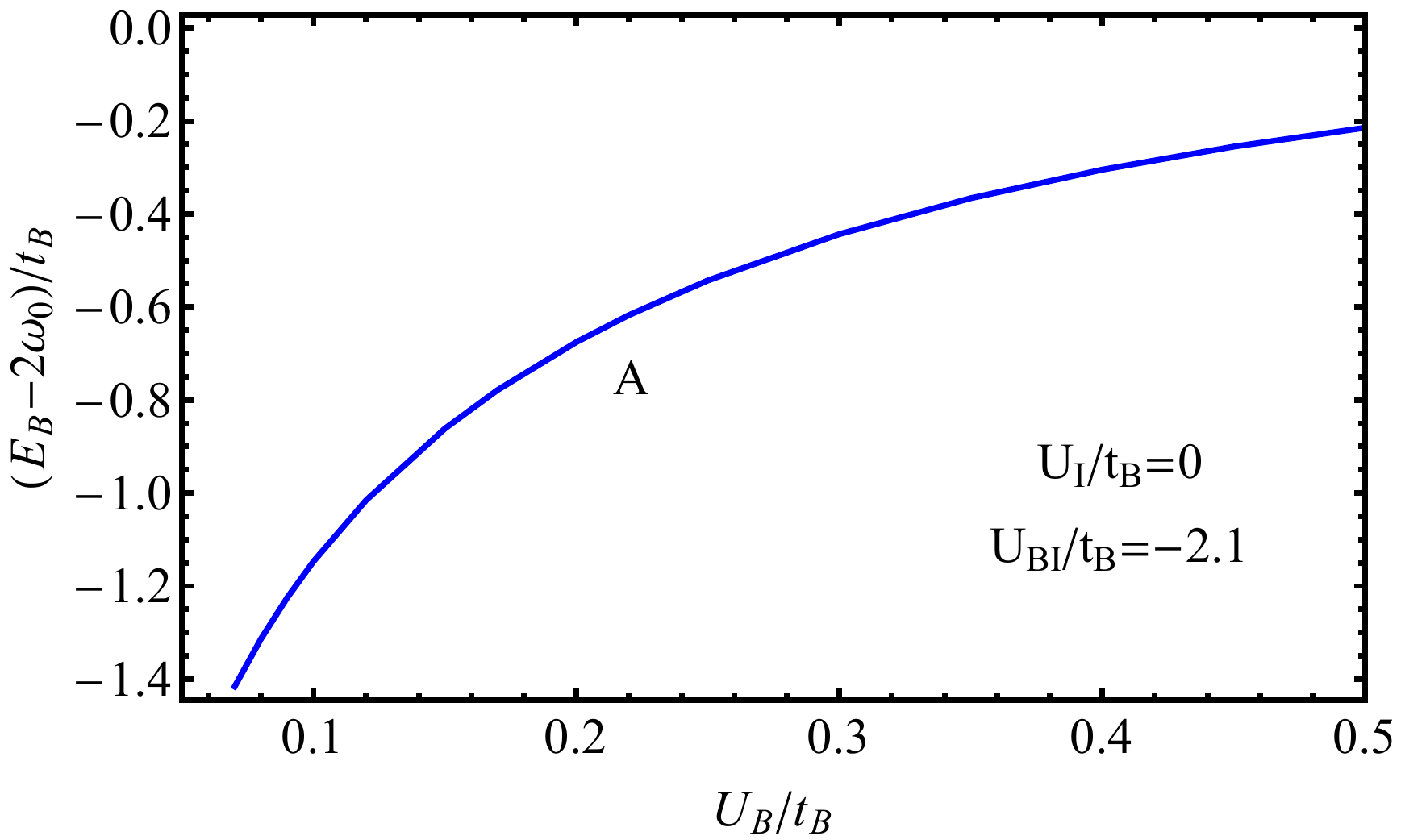}%
\caption{The top panel shows the energy of the ground state bipolaron (blue solid) 
relative to the
 continuum of two unbound attractive polarons with zero COM momentum as a function  of the impurity-impurity interaction strength $U_I$ for $U_{BI}/t_B=-2.1$.
  The dashed red line is the bound state
 energy of two impurities in an empty lattice. The bottom panel shows the energy of the ground state bipolaron as a function of the boson-boson
 repulsion $U_B$ for $U_{BI}/t_B=-2.1$ and $U_I=0$.}
\label{Fig:EnergiesUIandUB}
\end{figure}

To explore the effects of a non-zero direct interaction between the bare impurities, we plot in Fig.~\ref{Fig:EnergiesUIandUB} the energy of the ground state
bipolaron with A symmetry  as a function of
$U_I$ for   $U_{BI}/t_B=-2.1$. The dashed line shows the binding energy of the two impurities in an empty
lattice calculated from the  pole of Eq.~\eqref{ScatteringMatrix} with $U_{BI}$ replaced by $U_I$. As discussed in Sec.~\ref{TwobodySec}, there is always one
bound state below the continuum for $U_I< 0$ due to the 2D nature of the
lattice~\cite{Piil2007}. Figure \ref{Fig:EnergiesUIandUB} shows that the presence of the BEC
 increases the binding energy of the bipolaron for $U_{I}/t_B\gtrsim-11.7$ making it stable also for $U_I> 0$. This is  due to the attractive
 interaction mediated by the BEC.  Somewhat surprisingly however,  the bipolaron is less bound for $U_{I}/t_B\lesssim-11.7$.
 This is because the BEC decreases the impurity component
of the polarons as quantified through their residues $Z_{\mathbf k}$,
which in turn decreases the  attraction coming from the direct interaction between the bare impurities, as
can be seen explicitly  from the first term in Eq.~\eqref{QPint}.
Note that there are no excited dimer states with E symmetry in an empty lattice as is easily understood from the fact they have a node in the origin making them
 insensitive to an on-site interaction. Hence, a bipolaron containing two identical fermionic impurities is \emph{only} stable in the presence of a BEC.

 We finally explore how the binding energy of the bipolaron depends on the repulsion between the bosons in the BEC. In the lower panel of
 Fig.~\ref{Fig:EnergiesUIandUB}, the ground state bipolaron energy is plotted as a function
 of $U_B$ with  $U_I=0$, and $U_{BI}/t_B=-2.1$. The binding energy increases with decreasing boson-boson repulsion.
 This can be understood from the perturbative expression Eq.~\eqref{InducedPert} showing that the induced interaction increases with
 decreasing $U_B$. Physically it reflects that the BEC becomes more compressible with decreasing boson-boson repulsion making the induced interaction stronger.
One should however note that  our theory neglects retardation effects and therefore likely will break down as the speed of sound in the BEC decreases
for $U_B\rightarrow 0$. We expect retardation effects to decrease the binding energy of the bipolaron.


\section{Conclusions and Outlook}
We explored mobile impurities immersed in a BEC in a square lattice. By calculating the spectral properties of the impurities, we showed that the lattice gives rise to a new and stable polaron branch above the single particle continuum, which is not present for continuum systems. We
derived an expression for the induced interaction between two polarons mediated by the exchange of density oscillations in the BEC, which
takes into account strong impurity-boson correlations. Using this induced interaction in an effective Schr\"odinger equation for two
 polarons, we showed that it can be strong enough to bind two polarons into a bipolaron. The wave function of the ground state bipolaron was shown to be symmetric
 under particle exchange and therefore relevant for bosonic impurities, whereas the doubly degenerate first excited bipolaron states are odd making them relevant
 for fermionic impurities. We investigated the spatial correlations between the bosons and the impurity in the
polaron states, as well as between two impurities in the bipolaron states and showed that both should be observable using high resolution quantum gas microscopy
 in optical lattices. Our results show that optical lattices are a promising platform to explore new aspects of polaron physics such as their spatial
properties, and for  observing the elusive
bipolarons for the first time in quantum degenerate gases. This  would be a major breakthrough opening the door for quantum simulation experiments probing their
properties in detail.

This work  opens up several interesting research directions. Bipolarons can be considered as precursors of Cooper pairs in the limit of small impurity concentration, and a
fundamental question is therefore how the nature of bipolarons change with increasing impurity concentration. This is analogous to  the question concerning how the
high $T_c$ superconducting  phase emerges with increasing hole doping in the cuprates~\cite{SCALAPINO1995329,Lee2006}.
Another interesting problem concerns how the results presented in the present paper depend on temperature, and there are  different theoretical and experimental results
even for single polarons in continuum systems~\cite{Levinsen2017,Guenther2018,Dzsotjan2020,Field2020,Yan2020}.
Also, while the Fermi exclusion principle likely excludes bound states involving more fermionic impurities, a natural question is whether they exist
for bosonic impurities. We have here concentrated on small values of $U_B/t_B$ where the bosons are deep in the superfluid regime, and an
intriguing topic concerns the behaviour of the polarons and bi-polarons for larger  $U_B/t_B$ where the system undergoes a phase transition to a Mott insulating phase~\cite{Colussi2022}.
 Moreover, since a range of interesting systems with strong correlations, non-trivial topology,  and
quantum phase transitions have been realised in optical lattices~\cite{Bloch:2012aa,Gross2017}, an exciting new research direction is to use  impurities as a probe of many-body
physics~\cite{Grusdt:2016aa,Camacho-Guardian2019,Bouton2020,Edri2020,Alhyder2022}.
 Finally, we note that impurity dynamics in lattice systems is a topic of broad relevance also to  condensed matter physics.
Indeed, entirely new 2D Bose-Fermi lattice systems consisting of  excitons mixed with electrons are now created in
van der Waals bi-layers stacked with a relative twist angle~\cite{Andrei:2021aa}. In the limit of small electron concentration, this will realise lattice Bose polarons in a solid state setting.
Excitons have moreover been used as  impurities to detect Wigner crystals and Mott insulators in quantum materials~\cite{Smoleski2021,Shimazaki2021}.

\section*{Acknowledgements}
We acknowledge useful discussions with Zoe Yan. This work has been supported by the Danish National Research Foundation through the Center of Excellence ``CCQ'' (Grant agreement no.: DNRF156) (GMB, SD and AJ). G.A.D.-C. acknowledges a Consejo Nacional de Ciencia y Tecnolog\'ia (CONACYT) scholarship. A. C. G. acknowledges financial support from Grant UNAM DGAPA PAPIIT No. IN108620 and PAPIIT No. IA101923.


\begin{appendix}
\section{Derivation of the Pair-Propagator in vacuum}
\label{PairApp}
As discussed in the main text, the vacuum pair-propagator $\Pi_{\mathrm{v}}(\mathbf{0},\omega)$ (see Eq.\ref{pair}) can be evaluated analytically.  Here, we provide the main details of this derivation.
\begin{equation}
\Pi_{\mathrm{v}}(\mathbf{0},\omega) = \frac{1}{8\pi^2t_{B}}\int_{-\pi}^{\pi}dk_x\int_{-\pi}^{\pi}dk_y \frac{1}{z' + 2\cos k_x +2\cos k_y}
\label{DPP1}
\end{equation}
where $z'=(\omega-4t_B)/2t_B$.
Performing the integration with respect to $k_y$, we obtain the following result for $|z'|>4$:
\begin{equation}
\Pi_{\mathrm{v}}(\mathbf{0},\omega)= \frac{\text{sgn}(z')}{2\pi t_B}\int_{0}^{\pi}\frac{dk_x}{\sqrt{(z'+2\cos k_x)^{2}-4}}.
\label{DPP2}
\end{equation}
By expanding the quadratic term inside the root and introducing the new variable $s=\cos k_x$, it is straightforward to show that
\begin{equation}
\Pi_{\mathrm{v}}(\mathbf{0},\omega)= \frac{\text{sgn}(z')}{4\pi t_B}\int_{-1}^{1} \frac{ds}{\sqrt{(1-s)(1+s)(\lambda_{+}-s)(\lambda_{-}-s)}},
\label{DPP3}
\end{equation}
where $\lambda_{+} = 1-z'/2$ and $\lambda_{-}=-1-z'/2$. The resulting integral can be expressed in terms of the complete elliptic integral of the first kind $K(z')$ ~\cite{abramowitz1972handbook},
\begin{equation}
\Pi_{\mathrm{v}}(\mathbf{0},\omega)= \frac{1}{\pi z't_B}K\left(\frac{4}{|z'|}\right),
\label{DPP4}
\end{equation}
which gives the result in Eq. \ref{pair} for $|z'|>4$. To provide the expression for $|z'|<4$, we employ the following analytic continuation of the elliptic function:
\begin{equation}
K\left(\frac{4}{|z'|+i\eta}\right) = \frac{|z'|}{4}\left(K\left(\frac{|z'|}{4}\right)-iK\left(\sqrt{1-\frac{z'^2}{16}}\right)\right),
\label{DPP5}
\end{equation}
we then replace $z=z'/4$ to obtain Eq. \ref{pair} in the main text.
\section{Two-body bound state}\label{App-Sec-TB-Bound}
On a 2D square lattice, the two-body Schr\"odinger equation in real space with an on-site interaction is
\begin{equation}
\left[-\sum_{\substack{i=1,2\\ \sigma=x,y}}t_i\nabla^2_{i,\sigma}+U_{BI}\delta(\mathbf{r}_2-\mathbf{r}_1)\right]\Psi\left(\mathbf{r}_1,\mathbf{r}_2\right)=E\Psi\left(\mathbf{r}_1,\mathbf{r}_2\right)
\end{equation}
where $\nabla^2_{i,\sigma}\Psi\left(\mathbf{r}_i,\mathbf{r}_j\right)=\Psi\left(\mathbf{r}_i+\hat{\mathbf{e}}_{\sigma},\mathbf{r}_j\right)-2\Psi\left(\mathbf{r}_i,\mathbf{r}_j\right)+\Psi\left(\mathbf{r}_i-\hat{\mathbf{e}}_\sigma,\mathbf{r}_j\right)$ with $\hat{\mathbf{e}}_{\sigma}$ being the unit vector along the $\sigma$ direction.
Expanding the wave function in terms of the plane wave $\Psi\left(\mathbf{r}_1,\mathbf{r}_2\right)=\sum_{\mathbf{k}_1,\mathbf{k}_2}\Phi\left(\mathbf{k}_1,\mathbf{k}_2\right)\exp\left(i\mathbf{k}_1\mathbf{r}_1+i\mathbf{k}_2\mathbf{r}_2\right)$,
we obtain for the kinetic energy
\begin{equation}
\sum_{\mathbf{k}_1,\mathbf{k}_2}\sum_{\substack{i=1,2\\ \sigma=x,y}}\left[-2t_i\left(\cos k_{i\sigma}-1\right)\right]\Phi\left(\mathbf{k}_1,\mathbf{k}_2\right)e^{i\mathbf{k}_1\mathbf{r}_1+i\mathbf{k}_2\mathbf{r}_2}.
\label{App-Eq-K-k1k2}\end{equation}
Since the interaction $U_{BI}\delta(\mathbf{r}_2-\mathbf{r}_1)$ is only a function of the relative position $\mathbf{r}_2-\mathbf{r}_1$, the total momentum is a conserved quantity. Then it is convenient to describe the system in terms of total momentum $\mathbf{P}$ and relative momentum $\mathbf{k}$. Introducing the center-of-mass and the relative position as \cite{martikainen2008cooper}
\begin{equation}\begin{array}{cc}
\mathbf{R}=c\mathbf{r}_1+(1-c)\mathbf{r}_2 &\text{and} \quad \mathbf{r}=\mathbf{r}_1-\mathbf{r}_2
\end{array}\end{equation}
where $c$ is a coefficient giving the weight of the position of particle $1$ in the center of mass. Note that we are considering $t_{i,x}=t_{i,y}$ and consequently, $c_x=c_y=c$. Then the relation between $\mathbf{k}_1, \mathbf{k}_2$ and $\mathbf{P}, \mathbf{k}$ is
\begin{equation}\begin{cases}
\mathbf{P}c+\mathbf{k}=\mathbf{k}_1,\\
\mathbf{P}(1-c)-\mathbf{k}=\mathbf{k}_2.
\end{cases}
\end{equation}
We  obtain the condition for $c$
\begin{equation}
\frac{t_1}{t_2}=\frac{\sin\left[P_\sigma\left(1-c\right)\right]}{\sin\left(P_\sigma c\right)}.
\end{equation}
Then it is straightforward to derive that the kinetic energy in terms of $\mathbf{P}, \mathbf{k}$ for a specific $\mathbf{k}_1, \mathbf{k}_2$ is
\begin{equation}\begin{aligned}
K_{\mathbf{P},\mathbf{k}}=2\sum_{\sigma=x,y}\left[E_{P_\sigma}\left(\cos k_{\sigma}-1\right)+E_{P_\sigma}+t_1+t_2\right],
\end{aligned}\end{equation}
where $E_{P_\sigma}=-t_1\cos\left(P_\sigma c\right)-t_2\cos\left[P_\sigma\left(1-c\right)\right]$.

Since $\mathbf{P}$ is a conserved quantity, we fix $\mathbf{P}$ in the Schr\"odinger equation and project the equation to a specific relative momentum $\mathbf{k}$ state.
\begin{equation}\begin{aligned}
K_{\mathbf{P},\mathbf{k}}\varphi_\mathbf{k}
+\sum_{\mathbf{k}'}U_{BI}\varphi_{\mathbf{k}'}=E\varphi_\mathbf{k},
\end{aligned}\label{SE-contact}\end{equation}
where for simplicity we denote the wavefunction as $\varphi_\mathbf{k}$. Therefore,
\begin{equation}
\varphi_\mathbf{k}=\frac{-\sum_{\mathbf{k}'}U_{BI}\varphi_{\mathbf{k}'}}{K_{\mathbf{P},\mathbf{k}}-E}.
\end{equation}
After summing over $\mathbf{k}$ on both sides of the equation we obtain
\begin{equation}\begin{aligned}
\frac{1}{U_{BI}}
=\sum_\mathbf{k}\frac{-1}{K_{\mathbf{P},\mathbf{k}}-E}
=-\int d\epsilon \frac{N(\epsilon)}{\epsilon-E},
\end{aligned}\label{App-Eq-TB-Bound}\end{equation}
where $\epsilon=K_{\mathbf{P},\mathbf{k}}$ and $N(\epsilon)$ is the corresponding density of states.
For a fixed $\mathbf{P}$, $E_{P_\sigma}$ is a constant  and therefore, $N(\epsilon)$ is qualitatively the same as that for a single particle on a rectangular lattice \cite{pesz1986densities}, which is finite near the boundary of the continuum. Therefore, for any value of $U_{BI}<0$/$U_{BI}>0$ one can always find a two-body bound state with $E$ below/above the continuum to satisfy Eq.~\ref{App-Eq-TB-Bound}.

\section{Pair-Propagator in BEC}\label{App-Sec-Pi-BEC}
The impurity-boson pair-propagator in BEC is given by
\begin{equation}
\Pi\left(\mathbf{P},i\Omega\right)=-\frac{1}{\beta M}\sum_{i\omega,\mathbf{k}\in BZ}G_{11}\left(i\omega,\mathbf{k}\right)G_{I0}\left(i\Omega-i\omega,\mathbf{P}-\mathbf{k}\right)
\end{equation}
where $\beta=1/k_BT$ with $k_B$ being the Boltzmann constant and $T$ the temperature. $G_{I0}(i\omega,\mathbf{k})=1/(i\omega-\epsilon_{I\mathbf{k}})$ is the Green's function for bare impurity and the BEC normal Green's function is
\begin{equation}
G_{11}\left(i\omega,\mathbf{k}\right)=\frac{u_{\mathbf{k}}^2}{i\omega-E_{\mathbf{k}}}-\frac{v_{\mathbf{k}}^2}{i\omega+E_{\mathbf{k}}}
\end{equation}
with $v_{\mathbf{k}}^2=u_{\mathbf{k}}^2-1$.
It is straightforward to derive
\begin{equation}\begin{aligned}
\frac{1}{\beta}\sum_{i\omega}G_{11}\left(i\omega,\mathbf{k}\right)G_{I0}\left(i\Omega-i\omega,\mathbf{P}-\mathbf{k}\right)=&
\frac{u_\mathbf{k}^2 n_B(E_{\mathbf{k}})}{-i\Omega+E_{\mathbf{k}}+\epsilon_{I\mathbf{P}-\mathbf{k}}} -\frac{v_\mathbf{k}^2 n_B(-E_{\mathbf{k}})}{-i\Omega-E_{\mathbf{k}}+\epsilon_{I\mathbf{P}-\mathbf{k}}}\\
+n_B\left(i\Omega-\epsilon_{I\mathbf{P}-\mathbf{k}}\right)&\left(\frac{u_\mathbf{k}^2}{i\Omega-\epsilon_{I\mathbf{P}-\mathbf{k}}-E_{\mathbf{k}}}-\frac{v_\mathbf{k}^2}{i\Omega-\epsilon_{I\mathbf{P}-\mathbf{k}}+E_{\mathbf{k}}}\right)\\
\end{aligned}\label{Eq-GG-sumw}\end{equation}
where $n_B(x)=1/(\exp(\beta x)-1)$ is the bosonic distribution function.
For bosonic impurities $i\Omega=i2n\pi/\beta$ ( with $n$ an integer), $n_B\left(i\Omega-\epsilon_{I\mathbf{P}-\mathbf{k}}\right)=n_B\left(-\epsilon_{I\mathbf{P}-\mathbf{k}}\right)$. For a single impurity, the distribution function of the impurity vanishes, thus $n_B\left(i\Omega-\epsilon_{I\mathbf{P}-\mathbf{k}}\right)=-1$
and at $T=0$, the pair-propagator is
\begin{equation}
\Pi\left(\mathbf{P},i\Omega\right)=\frac{1}{M}\sum_{\mathbf k}\frac{u_\mathbf{k}^2}{i\Omega-\epsilon_{I\mathbf{P}-\mathbf{k}}-E_{\mathbf{k}}}.
\label{Eq-sum-iomega-T0}\end{equation}
For fermionic impurities the Matsubara frequencies $i\Omega=i(2n+1)\pi/\beta$, which yields
$n_B\left(i\Omega-\epsilon_{I\mathbf{P}-\mathbf{k}}\right)=-f_F\left(-\epsilon_{I\mathbf{P}-\mathbf{k}}\right)$
with the Fermi distribution function $f_F(x)=1/(\exp(\beta x)+1)$. Considering the limit of a single impurity, one still has $n_B\left(i\Omega-\epsilon_{I\mathbf{P}-\mathbf{k}}\right)=-1$ and at $T=0$, Eq. \ref{Eq-sum-iomega-T0}.
Therefore, one ends up with Eq. \ref{Tmatrix} for the pair propagator in the main text.
\section{Spatial Correlations, Wavefunction and $\mathcal{T}$-matrix}
\label{SinglePolaronAppendix}
{\it Spatial correlations.-} Let us provide further details on the spatial properties of the polaron, such features are obtained from a many-body polaron wavefunction equivalent to the $\mathcal{T}$-matrix formalism. Let us start discussing the spatial correlations given by Eq. \ref{CorrFn} in the main text.
In Eq. \ref{CorrFn}, $\hat n_B(\mathbf i)=\hat b_{\mathbf{i}}^\dagger\hat b_{\mathbf{i}},$ which we conveniently write in momentum space to account for the condensed atoms forming the BEC, thus, we have
\begin{equation}\begin{aligned}
\hat n_B(\mathbf i)=&n_0+
\frac{\sqrt{n_0}}{\sqrt{M}}\sum_{\mathbf q\neq0}\left(e^{-i\mathbf q\cdot \mathbf i}\hat b^\dagger_{\mathbf q}+e^{i\mathbf q\cdot \mathbf i}\hat b_{\mathbf q}\right)+\frac{1}{M}\sum_{\substack{\mathbf k\neq0\\ \mathbf q\neq\mathbf{k}}}  e^{-i\mathbf q\cdot \mathbf i}\hat b^\dagger_{\mathbf k}\hat b_{\mathbf k-\mathbf q}\\
=&n_0+n_{\mathrm{\mathrm{ex}}}+
\frac{\sqrt{n_0}}{\sqrt{M}}\sum_{\mathbf q}(u_{\mathbf q}+v_\mathbf q)\left(e^{-i\mathbf q\cdot \mathbf i}\hat \beta^\dagger_{\mathbf q}+e^{i\mathbf q\cdot \mathbf i}\hat \beta_{\mathbf q}\right)\\
&+\frac{1}{M}\sum_{\mathbf k,\mathbf q} ( e^{-i\mathbf q\cdot \mathbf i}\hat \beta^\dagger_{\mathbf k}\hat \beta_{\mathbf k-\mathbf q}u_{\mathbf k}u_{\mathbf k-\mathbf q}+ e^{i\mathbf q\cdot \mathbf i}\beta^\dagger_{\mathbf k-\mathbf q}\hat \beta_{\mathbf k}v_{\mathbf k}v_{\mathbf k-\mathbf q})\\
&+\frac{1}{M}\sum_{\mathbf k,\mathbf q}  e^{-i\mathbf q\cdot \mathbf i}(\hat \beta^\dagger_{\mathbf k}\hat \beta^\dagger_{\mathbf q-\mathbf k}u_{\mathbf k}v_{\mathbf q-\mathbf k}+\hat \beta_{-\mathbf k}\hat \beta_{\mathbf k-\mathbf q}v_{-\mathbf k}u_{\mathbf k-\mathbf q}),
\end{aligned}\end{equation}
written in terms of the standard Bogoliubov operators
\begin{subequations}
\begin{equation}
\hat b_{\mathbf k}=u_{\mathbf k}\hat \beta_{\mathbf k}+v_{-\mathbf k}\hat \beta^\dagger_{-\mathbf k},
\end{equation}
\begin{equation}
\hat b^\dagger_{\mathbf k}=u_{\mathbf k}\hat \beta^\dagger_{\mathbf k}+v_{-\mathbf k}\hat \beta_{-\mathbf k}.
\end{equation}
\end{subequations}
$n_{\mathrm{ex}}=\sum_{\mathbf k\neq0}v_{\mathbf k}^2/M$ is the density of the depletion of the ground state BEC due to finite $U_B$.
To evaluate the correlation function we employ the many-body polaron wavefunction defined by Eq. \ref{Chevy} in the main text and take the approximation $n=\langle \Psi_P| \hat n_B({\mathbf i})|\Psi_P\rangle=\langle BEC| \hat n_B({\mathbf i})|BEC\rangle=n_0+n_{\mathrm{ex}}=n_0$.
After some algebra we obtain
 \begin{equation}\begin{aligned}
\langle \Psi_P| [\hat n_B(\mathbf i)-n]\hat{n}_I(\mathbf 0)|\Psi_P\rangle
=&\frac{\sqrt{n}}{M^{3/2}}\sum_{\mathbf p}(u_{-\mathbf p}+v_{-\mathbf p})\left(\phi_0^*\psi_{\mathbf p}e^{-i\mathbf p\cdot\mathbf i}+\phi_0\psi^*_{\mathbf p}e^{i\mathbf p\cdot\mathbf i}\right)\\
&+\frac{1}{M^2}\sum_{\mathbf p,\mathbf p'}e^{-i(\mathbf p-\mathbf p')\cdot\mathbf i}\psi^*_{\mathbf p'}\psi_{\mathbf p}(u_{-\mathbf p}u_{-\mathbf p'}+v_{\mathbf p}v_{\mathbf p'}).
\end{aligned}\end{equation}
This expression depends on the properties of the BEC as well as the  parameters of the many-body wave function $\phi_0$ and $\psi_{\mathbf k}.$ We make use of the properties $u_{\mathbf k}=u_{-\mathbf k},$  and  $v_{\mathbf k}=v_{-\mathbf k},$  as well as the wavefunction follows $\psi_{\mathbf k}=\psi_{-\mathbf k},$ as we will explicitly show.

{\it Many-body wavefunction.-} To determine the parameters of Eq.~\ref{Chevy} we employ a variational approach which as we will illustrate is equivalent to the field theory employed with the self-energy as in Eq.~\ref{SigmaP}. We start from the energy functional for $\omega_{\mathbf 0}$
\begin{equation}\begin{aligned}
&\omega_{\mathbf 0}\left(|\phi_0|^2+\sum_{\mathbf k}|\psi_{\mathbf k}|^2\right)
=\sum_{\mathbf k}\left(\epsilon_{I\mathbf k}+E_{\mathbf k}\right)|\psi_{\mathbf k}|^2+\epsilon_{I\mathbf 0}|\phi_0|^2+nU_{BI}|\phi_0|^2\\
&+\frac{U_{BI}\sqrt{n_0}}{\sqrt{M}}\sum_{\mathbf k}(u_{\mathbf k}+v_{\mathbf k})\left(\phi_0\psi_\mathbf k^*+\phi_0^*\psi_{\mathbf k}\right)
+\frac{U_{BI}}{M}\sum_{\mathbf p,\mathbf p'}(u_{-\mathbf p}u_{-\mathbf p'}+v_{\mathbf p'}v_{\mathbf p})\psi^*_{\mathbf p'}\psi_{\mathbf p},
\end{aligned}\end{equation}
where  $\phi_0$ and $\phi_0^*$ as well as $\psi_{\mathbf k}$ and $\psi_{\mathbf k}^*$  are treated as independent parameters. Then, by minimizing the energy with respect to $\phi_0^*$ and  $\psi_{\mathbf k}^*$  we obtain the following set of equations
\begin{subequations}
\begin{equation}
0=\frac{U_{BI}\sqrt{n}}{\sqrt{M}}\sum_{\mathbf k}u_{\mathbf k}\psi_{\mathbf k}-(\omega_{\mathbf 0}-\epsilon_{I\mathbf 0}-nU_{BI})\phi_0,
\label{Eqv1}\end{equation}
\begin{equation}\begin{aligned}
0=\left(\epsilon_{I\mathbf k}+E_{\mathbf k}-\omega_{\mathbf 0}\right)\psi_{\mathbf k}
+\frac{U_{BI}\sqrt{n}}{\sqrt{M}}u_{\mathbf k}\phi_0+u_{\mathbf k}\frac{U_{BI}}{M}\sum_{\mathbf p} u_{\mathbf p}\psi_{\mathbf p},
\end{aligned}\label{Eqv2}\end{equation}
\end{subequations}
where for consistency with our Green's function formalism we only retain the $u_{\mathbf k}$ coherence factors and take $n_0=n$.

To determine the amplitude of the wavefunction in terms of the quasiparticle properties we have from the Eq. \ref{Eqv1}
\begin{equation}
\sum_{\mathbf k}u_{\mathbf k}\psi_{\mathbf k}=\frac{\sqrt{M}}{U_{BI}\sqrt{n}}(\omega_{\mathbf 0}-\epsilon_{I\mathbf 0}-nU_{BI})\phi_0,
\end{equation}
which can be replaced into the Eq. \ref{Eqv2}. Thus, if the polaron energy and its quasiparticle residue are known, then we can directly determine the many-body wave function with
\begin{equation}
\label{psi}
\psi_{\mathbf k}=-\frac{\omega_{\mathbf 0}-\epsilon_{I\mathbf 0}}{\epsilon_{I\mathbf k}+E_{\mathbf k}-\omega_{\mathbf 0}}\frac{u_{\mathbf k} \sqrt{Z_0}}{\sqrt{M n}},
\end{equation}
where the amplitude $\phi_0=\sqrt{Z_0}$ has been applied.

{\it Wavefunction and the $\mathcal{T}$-matrix.-} To obtain the polaron wavefunction we employ the quasiparticle properties from the $\mathcal{T}$-matrix formalism, this is justified by the equivalence between these two approaches as we will briefly discuss.  To prove the equivalence between these approaches, let us introduce the following parameter,
\begin{gather}
\chi=\phi_0+\sum_{\mathbf k}u_{\mathbf k}\psi_{\mathbf k}.
\label{chi}
\end{gather}
Writing $\phi_0$ in terms of $\chi$ using Eq.~\ref{Eqv1} we obtain
\begin{gather}
\phi_0=U_{BI}\sqrt{\frac{n}{M}}\frac{\chi}{\omega_{\mathbf 0}-\epsilon_{I\mathbf 0}-nU_{BI}+U_{BI}\sqrt{\frac{n}{M}}}.
\label{phi01}
\end{gather}
Further replace this expression into Eq.~\ref{Eqv2} to write $\psi_{\mathbf k}$ in terms of $\chi$
\begin{equation}\begin{aligned}
\psi_{\mathbf k}=\frac{u_{\mathbf k}}{\omega_{\mathbf 0}-\epsilon_{I\mathbf k}-E_{\mathbf k}}\frac{U_{BI}}{M}
\frac{\left(\omega_{\mathbf 0}-\epsilon_{I\mathbf 0}\right)\chi}{\omega_{\mathbf 0}-\epsilon_{I\mathbf 0}-nU_{BI}+U_{BI}\sqrt{\frac{n}{M}}}.
\end{aligned}\label{psi1}
\end{equation}
Now, we replace Eq.~\ref{phi01} and Eq.~\ref{psi1} in the expression for $\chi$ in Eq.~\ref{chi}
\begin{equation}\begin{aligned}
\chi=\frac{\chi}{\omega_{\mathbf 0}-\epsilon_{I\mathbf 0}-nU_{BI}+U_{BI}\sqrt{\frac{n}{M}}}
\left[U_{BI}\sqrt{\frac{n}{M}}+\frac{U_{BI}}{M}\sum_{\mathbf k}\frac{u_{\mathbf k}^2\left(\omega_{\mathbf 0}-\epsilon_{I\mathbf 0}\right)}{\omega_{\mathbf 0}-\epsilon_{I\mathbf k}-E_{\mathbf k}}\right].
\end{aligned}\end{equation}
After some algebra we arrive at a self-consistent equation for $\omega_{\mathbf 0}$
 \begin{gather}
 \omega_{\mathbf 0}-\epsilon_{I\mathbf 0}=\frac{nU_{BI}}{1-\frac{U_{BI}}{M}\sum_{\mathbf k}\frac{u^2_{\mathbf k}}{\omega_{\mathbf 0}-\epsilon_{I\mathbf k}-E_{\mathbf k}}}=n\mathcal T(\omega_{\mathbf 0}),
 \end{gather}
 thus the quasiparticle energy of the polaron is given by
 \begin{gather}
  \omega_{\mathbf 0}=\epsilon_{I \mathbf 0}+n\mathcal T(\omega_{\mathbf 0}),
 \end{gather}
 which is precisely the equation for the polaron energy employed within the Green's function formalism.
\section{Induced interaction between polarons}\label{App-Sec-Vind}
\begin{figure}[h]
\centering
\includegraphics[width=0.7\textwidth]{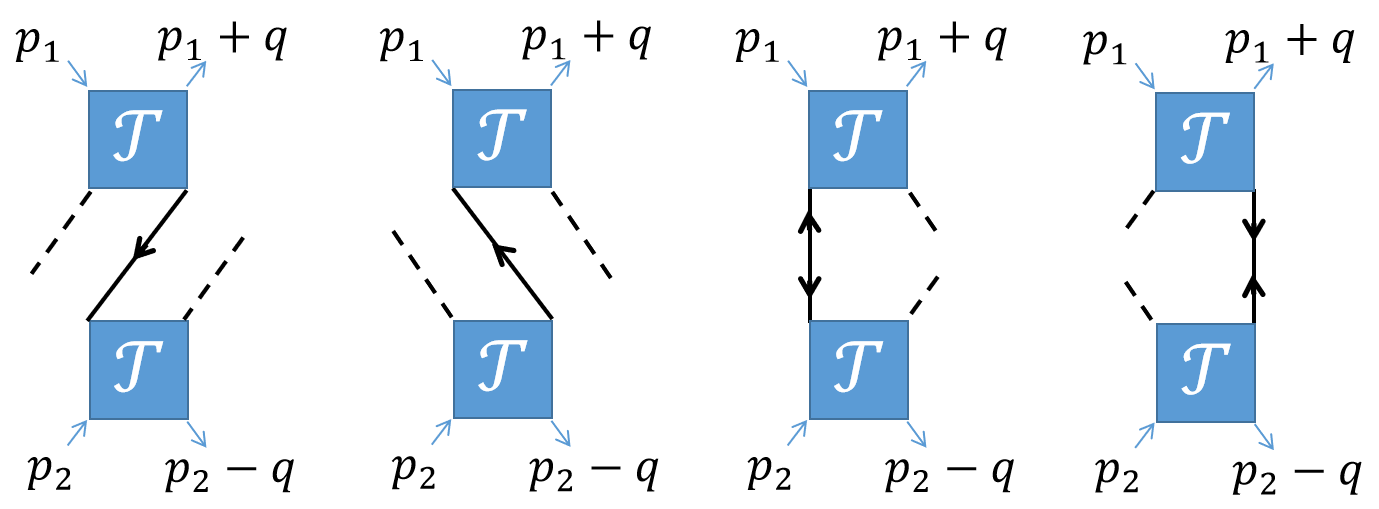}%
\caption{The diagram for the induced interaction between two impurities due to the exchange of the Bogoliubov mode. The solid black lines are the normal and anomalous Bogoliubov Green's functions for the bosons, dashed lines correspond to condensate particles. The boson-impurity interaction is represented by the blue boxes and is taken within the ladder approximation.}
\label{Fig-Vind-diagram}
\end{figure}

The induced interaction between two impurities mediated by a BEC is given by the diagram shown in Figure. \ref{Fig-Vind-diagram}.
The analytical expression is
\begin{equation}\begin{aligned}
V^I_{\mathrm{ind}}\left(p_1,p_2,q\right)=n_0
&[\mathcal{T}\left(p_1\right)G_{11}\left(-q\right)\mathcal{T}\left(p_2-q\right)+\mathcal{T}\left(p_1+q\right)G_{11}\left(q\right)\mathcal{T}\left(p_2\right)\\
&+\mathcal{T}\left(p_1+q\right)G_{12}\left(q\right)\mathcal{T}\left(p_2-q\right)+\mathcal{T}\left(p_1\right)G_{21}\left(q\right)\mathcal{T}\left(p_2\right)]
\end{aligned}\end{equation}
where $p_1, p_2, q$ are four momentums and the anomalous Green's functions of BEC
\begin{equation}
G_{12}\left(i\omega,\mathbf{k}\right)=G_{21}\left(i\omega,\mathbf{k}\right)=u_{\mathbf{k}}v_{\mathbf{k}}\left(\frac{1}{i\omega-E_{\mathbf{k}}}-\frac{1}{i\omega+E_{\mathbf{k}}}\right)
\end{equation}
with $u_{\mathbf{k}}v_{\mathbf{k}}=-n_0U_{B}/2E_{\mathbf{k}}$.
Suppose $\tilde{E}_B$ is the binding energy of the produced bipolaron and $m^*$ the effective mass of the polaron. If $\sqrt{|\tilde{E}_B|/m^*}$ is far less than the sound velocity of the BEC, the time consumed by the density wave in BEC to transport the induced interaction between polarons can be ignored. Therefore, $V^I_{\mathrm{ind}}(p_1,p_2,q)$ is approximately a frequency transfer $iq_0$ independent function. In principal, we can take any value for $iq_0$ in $V^I_{\mathrm{ind}}$ and in our calculation, we take $iq_0=0$.
Since the variations of the energies of polarons during scattering should be of order of $\tilde{E}_B$, one can approximately treat the energy components of $p_1$ and $p_2$ in $V^I_{\mathrm{ind}}$ as appropriate constants if $\tilde{E}_B$ is far less than the typical energy of the Bogoliubov excitation.
As a result, the induced interaction between a pair of polarons with total momentum being $0$ is given by
\begin{equation}\begin{aligned}
V_{\mathrm{ind}}\left(\mathbf{k},\mathbf{k}'\right)=Z_{\mathbf{k}}^2n_0[&\mathcal{T}\left(\omega,\mathbf{k}\right)G_{11}\left(0,\mathbf{k}-\mathbf{k}'\right)\mathcal{T}\left(\omega,-\mathbf{k}'\right)
+\mathcal{T}\left(\omega,\mathbf{k}'\right)G_{11}\left(0,\mathbf{k}'-\mathbf{k}\right)\mathcal{T}\left(\omega,-\mathbf{k}\right)\\
+&\mathcal{T}\left(\omega,\mathbf{k}'\right)G_{12}\left(0,\mathbf{k}'-\mathbf{k}\right)\mathcal{T}\left(\omega,-\mathbf{k}'\right)
+\mathcal{T}\left(\omega,\mathbf{k}\right)G_{21}\left(0,\mathbf{k}'-\mathbf{k}\right)\mathcal{T}\left(\omega,-\mathbf{k}\right)]
\end{aligned}\end{equation}
where $\mathbf{k}$ and $\mathbf{k}'$ are respectively the relative momentums before and after scattering.
Considering
\begin{subequations}
\begin{equation}
G_{11}\left(i\omega,\mathbf{k}\right)=G_{11}\left(i\omega,-\mathbf{k}\right)
\end{equation}
\begin{equation}
\mathcal{T}\left(i\omega,\mathbf{k}\right)=\mathcal{T}\left(i\omega,-\mathbf{k}\right)
\end{equation}
\end{subequations}
one has the Eq. \ref{InducedInt} in the main text.

Now $V(p_1,p_2;q_1)$ can be taken out from the frequency summation in Eq. \ref{Eq-BS}. If the pole expansion for impurity Green's function $G_I(i\omega, \mathbf{k})\simeq Z_{\mathbf{k}}/(i\omega-\omega_{\mathbf{k}})$ is taken, this summation can be calculated analytically,
\begin{equation}\begin{aligned}
\frac{1}{\beta}\sum_{iq_{0}}G_I\left(iq_{0},\mathbf{k}\right)G_I\left(i\Omega-iq_{0},-\mathbf{k}\right)
=-\frac{Z_{\mathbf{k}}^2}{i\Omega-2\omega_{\mathbf{k}}}
\end{aligned}\end{equation}
which only depends on the total frequency $i \Omega$ of the two polarons.
As a result, the Bethe-Salpeter equation reduces to be the Lippmann-Schwinger equation
\begin{equation}\begin{aligned}
\Gamma_{\mathrm{eff}}\left(\mathbf{k},\mathbf{k}',i\Omega\right)=V_{\mathrm{qp}}\left(\mathbf{k},\mathbf{k}'\right)
+\frac{1}{ M}\sum_{\mathbf{q}_1}
\frac{V_{\mathrm{qp}}\left(\mathbf{k},\mathbf{k}+\mathbf{q}_1\right)}{i\Omega-2\omega_{\mathbf{k}+\mathbf{q}_1}}\Gamma_{\mathrm{eff}}\left(\mathbf{k}+\mathbf{q}_1,\mathbf{k}',i\Omega\right)
\end{aligned}\label{Eq-Gamma-eff}\end{equation}
where we have changed the notation so that the three arguments $\mathbf{k},\mathbf{k}'$ and $i\Omega$ in $\Gamma_{\mathrm{eff}}$ respectively represent the relative momentum of incoming and outgoing polarons as well as the total frequency of the two polarons. Eq. \ref{Eq-Gamma-eff} is equivalent to the Schr\"odinger equation Eq. \ref{Eq-eff-SE}.
\end{appendix}



\bibliography{SciPost-Ref-LP}

\nolinenumbers

\end{document}